\documentclass[longauth]{aa} % if the references are not structured 
\usepackage{graphicx}
\usepackage{multicol}
\usepackage{booktabs}
\usepackage{natbib}
\usepackage{multirow}
\newcommand\Tstrut{\rule{0pt}{2.6ex}}         % = `top' strut
\newcommand\Bstrut{\rule[-0.9ex]{0pt}{0pt}}   % = `bottom' strut
\bibpunct{(}{)}{;}{a}{}{,} % to follow the A&A style
\usepackage{xargs}
\usepackage{amsmath}
\usepackage{longtable}
\usepackage{array}
\newlength\mylength
\newcolumntype{C}[1]{>{\centering\arraybackslash}p{#1}}
\usepackage{tabularx}
\usepackage{txfonts}
\usepackage{verbatim}
\usepackage[linecolor=blue,backgroundcolor=blue!25,bordercolor=blue,obeyFinal]{todonotes} 
\usepackage{hyperref}
\usepackage{lscape, afterpage}
\hypersetup{
    colorlinks = true,
    citecolor = {blue},
    linkcolor = {blue},
    linkbordercolor = {white}}

\usepackage[normalem]{ulem}

\def\13co{$^{13}$CO}
\def\c18o{C$^{18}$O}
\def\kms{km~s$^{-1}$}

\begin{document} 
   \authorrunning{Suri et al.}
   \titlerunning{The CARMA-NRO Orion Survey: The filamentary structure as seen in C$^{18}$O emission}
   \title{The CARMA-NRO Orion Survey}
   \subtitle{The filamentary structure as seen in C$^{18}$O emission}

   \author{S\"umeyye T.~Suri%\thanks{text of footnote}
          \inst{1}\fnmsep\thanks{Current address: Max Planck Institute for Astronomy, K\"onigstuhl 17, 69117 Heidelberg, Germany},
          %\inst{2},
          \'Alvaro S\'anchez-Monge
          \inst{1},
          Peter Schilke
          \inst{1},
          Seamus D.~Clarke
          \inst{1},
          Rowan J.~Smith
          \inst{3},\\
          Volker Ossenkopf-Okada
          \inst{1},
          Ralf Klessen
          \inst{4},
          Paolo Padoan
          \inst{5}$^,$
          \inst{6},
          Paul Goldsmith
          \inst{7},
           H\'ector G.~Arce
          \inst{8},
          John Bally
          \inst{9},\\
          John M. Carpenter
          \inst{10},
          Adam Ginsburg
          \inst{11}, 
          Doug Johnstone
          \inst{12}$^,$
          \inst{13},
          Jens Kauffmann
          \inst{14},
          Shuo Kong
          \inst{8},
          Dariusz C. Lis
          \inst{15}$^,$
          \inst{16},\\
          Steve Mairs
          \inst{17},
          Thushara Pillai
          \inst{18},
          Jaime E. Pineda
          \inst{19},
          Ana Duarte-Cabral
          \inst{20}
          }
          
   \institute{I.~Physikalisches Institut, Universit\"at zu K\"oln,
              Z\"ulpicher Str. 77, D-50937 K\"oln, Germany\label{inst1}\\
            \email{suri@mpia.de}
            \and Jodrell Bank Centre for Astrophysics, School of Physics and Astronomy, University of Manchester, Oxford Road, Manchester M13 9PL, UK\label{inst2} 
            \and Universit\"{a}t Heidelberg, Zentrum f\"{u}r Astronomie, Albert-Ueberle-Str. 2, 69120 Heidelberg, Germany \label{inst3}
            \and Institut de Ci\`{e}ncies del Cosmos, Universitat de Barcelona, IEEC-UB, Mart\'i i Franqu\`es 1, E08028 Barcelona, Spain; ppadoan@icc.ub.edu \label{inst4}
            \and ICREA, Pg. Llu\'{i}s Companys 23, 08010 Barcelona, Spain\label{inst5}
            \and Jet Propulsion Laboratory, California Institute of Technology, 4800 Oak Grove Drive, Pasadena, CA 91109, USA\label{inst6} 
   			\and Department of Astronomy, Yale University, New Haven, Connecticut 06511, USA \label{inst7}
            \and Department of Astrophysical and Planetary Sciences, University of Colorado, Boulder, Colorado, USA \label{inst8}
            \and Joint ALMA Observatory, Alonso de C\'ordova 3107 Vitacura, Santiago, Chile \label{inst9}
            \and National Radio Astronomy Observatory,  1003 Lopezville road, Socorro NM, 87801\label{inst10}
            \and NRC Herzberg Astronomy and Astrophysics, 5071 West Saanich Rd, Victoria, BC, V9E 2E7, Canada \label{inst11}
            \and Department of Physics and Astronomy, University of Victoria, Victoria, BC, V8P 5C2, Canada \label{12}
            \and Haystack Observatory, Massachusetts Institute of Technology, 99 Millstone Road, Westford, MA 01886, USA\label{inst13}     
            \and Sorbonne Universit\'e, Observatoire de Paris, Universit\'e PSL, CNRS, LERMA, F-75014, Paris, France\label{inst14}
            \and Cahill Center for Astronomy and Astrophysics 301-17, California Institute of Technology, Pasadena, CA 91125, USA \label{inst15}
            \and East Asian Observatory, 660 N. A'ohoku Place, Hilo, Hawaii, 96720, USA;  s.mairs@eaobservatory.org\label{inst16}
            \and Max--Planck--Institut f\"ur Radioastronomie, Auf dem H\"ugel 69, D--53121 Bonn, Germany\label{17}
            \and Max-Planck-Institut f\"ur extraterrestrische Physik, Giessenbachstrasse 1, 85748 Garching, Germany \label{inst18}	
            \and School of Physics and Astronomy, Cardiff University, Cardiff, CF24 3AA, UK\label{19}
            }
  \date{Received 08, 2018; accepted 12, 2018}
  \abstract
   {We present an initial overview of the filamentary structure in the Orion A molecular cloud utilizing a high angular and velocity resolution \c18o(1--0) emission map that was recently produced as part of the CARMA-NRO Orion Survey.} % AG suggests change
   {The main goal of this study is to build a credible method to study varying widths of filaments which has previously been linked to star formation in molecular clouds. Due to the diverse star forming activities taking place throughout its $\sim$20~pc length, together with its proximity of 388~pc, the Orion A molecular cloud provides an %AG
   excellent laboratory for such an experiment to be carried out with high resolution and high sensitivity.}
   {Using the widely-known structure identification algorithm, DisPerSE,  on a 3-dimensional (PPV) \c18o cube, we identified 625 relatively short (the longest being 1.74~pc) filaments over the entire cloud. We study the distribution of filament widths using \texttt{FilChaP}, a \texttt{python} package that we have developed and made publicly available.} %AG
   {We find that the filaments identified in  a 2 square degree PPV cube do not overlap spatially, except for the complex OMC-4 region that shows distinct velocity components along the line of sight. The filament widths vary between 0.02 and 0.3 pc depending on the amount of substructure that a filament possesses. The more substructure a filament has, the larger is its width. We also find that despite this variation, the filament width shows no anticorrelation with the central column density which is in agreement with previous \textit{Herschel} observations.}
   {}


   \maketitle
%
%________________________________________________________________

\section{Introduction}
Over the past decade, it has become clear that stars preferentially form along and at the junctions of filaments inside cold, dense molecular clouds \citep[e.g.][]{Schneider2012}. Filaments pervade molecular clouds whether these clouds form high-mass or low-mass stars, or lack star formation entirely \citep{Andre2010,Andre2016}. Earlier studies of dust continuum observed with the \textit{Herschel} Space Observatory suggest a ``universal width'' of the observed filaments \citep{Arzoumanian2011}. The distribution of widths of 90 filaments having column densities of 10$^{20}$ -- 10$^{23}$~cm$^{-2}$ peaks at 0.1~pc with very little dispersion ($\pm$0.03~pc, \citealt{Arzoumanian2011}). However, work by \citet{Smith2014} shows that the calculated filament width is dependent on the fitting range considered when analyzing the radial profiles. They found that the degeneracy between the central density of the filaments and their inner flat-radius raises uncertainties. We also refer to \citet{Pano2017} for the shortcomings of fitting methods used in calculating filament widths. Moreover, higher angular resolution observations that allow study of high density filaments within molecular clouds suggest widths as narrow as 0.01~pc \citep{Sanchez2014,Hacar2018}, while observations of tracers such as \13co have revealed filaments with 0.4~pc widths \citep{Pano2014}. For the study of filament properties, in particular their widths, molecular line observations allow studying intertwined filamentary networks which otherwise appear as single objects when observed in dust emission due to projection of 3-dimensional information (position-position-velocity, hereafter; PPV) on to 2-dimensional column density space, therefore providing a more accurate determination of the filament properties.
Understanding the role of filaments in the process of star formation requires high spatial and spectral resolution studies of, ideally, multiple molecular emission lines that trace different densities within the cloud, as well as continuum emission. This way, filament properties can be calculated for different tracers that probe a variety of physical and chemical conditions. Investigating physical and kinematic  properties of filaments weighted by the environmental and chemical conditions allows us to gain an understanding of how filaments form, how this relates to the formation of molecular clouds, how mass flow and/or accretion onto filaments leads to the formation of new stars (see e.g.~\citealt{Ostriker1964,Larson1985,Inutsuka1992,Pon12,Fischera12,Heitsch2013b,Smith2014,Clarke15,Seifried15,Clarke16,Smith2016,Clarke17,Chira2017}).

In this work, we study filament properties towards the closest site of ongoing high-mass star formation, the Orion A molecular cloud \citep[$d\approx 388$\,pc,][]{Kounkel2017}. 
Orion A is amongst the most studied regions in the Galaxy \citep[e.g.][]{Bally1987,Johnstone1999,Shimajiri2009,Nakamura2012}. 
The northern Orion A region, encompassing OMC-1, OMC-2, OMC-3 and OMC-4, is commonly referred as the Integral Shaped Filament (hereafter; ISF, \citealt{Bally1987}). The whole ISF is subject to ionizing UV radiation produced by the Trapezium stars 
%at the heart of the filament, 
$\sim1$\,pc in front of the molecular cloud \citep{vanderWerf2013}. 
Moreover, the NGC1977 H{\sc{II}} region, powered by a B0V type star, 42 Ori, could produce a radiation field as strong as 2000 times the standard  value of the ISRF (G$_0$ $\simeq$ 2000) on the OMC-3 filaments, depending on the 3D orientation of the star relative to the northern ISF. Towards the south, the ISF extends to the L1641 region that is more diffuse compared to the northern portion, and less affected by the UV radiation due to the lack of high-mass stars.The structures seen in Orion A are therefore shaped not only by the gravitational potential of OMC-1, but also by the feedback from high-mass and low-mass (proto)stars.

So far, large-scale spectral line emission maps of the Orion A cloud have been obtained only with low angular resolution, while high angular resolution observations have been obtained only towards selected sub-regions of the cloud. For example, while \citet{Bally1987} observed a 1.5~square degree region of the ISF
with a 0.1~pc resolution, \citet{Kainulainen2017} has observed a 3 x 11 arcmin (0.33 x 1.24 pc) region towards OMC-2 with a $\sim$1200 AU resolution. 
Similarly, \citet{Hacar2018} observed a narrow field around the OMC-1/OMC-2 regions with high angular resolution (1750 AU) using a map combining data from ALMA and the IRAM 30m telescope. 
The current work, on the other hand, provides a first look into filament properties that are obtained with high angular resolution in an extensive map of the Orion A molecular cloud.

In Section~\ref{sec:observations}, we summarize the large scale and high angular resolution observations from the CARMA-NRO Orion Survey. In Section~\ref{sec:filamentProperties} we introduce our methods to calculate physical properties of filaments. Based on these methods we present filament properties as observed in \c18o emission, see Section~\ref{sec:discussion}. Finally, we summarize our results in Section~\ref{sec:conclusions}.

\section{Observations}\label{sec:observations}

In order to image an extended region towards Orion A with high angular resolution and high sensitivity, the CARMA-NRO Orion Survey (PI: J.~Carpenter) was granted $\sim$650 hours of observing time at the Combined Array for Millimeter-Wave Astronomy (CARMA). The observations were carried out between 2013 and 2015 using 15 antennas (6 and 10~m dishes) of the array with D and E configurations resulting in an angular resolution of about 8~arcsec (0.015~pc at a distance of 388~pc). Outflow and diffuse gas tracers ($^{12}$CO and $^{13}$CO), warm dense gas tracers (CS and C$^{18}$O), a cold dense gas tracer (CN) and a shock tracer (SO) along with the 3 mm continuum were observed simultaneously. 
The interferometric observations were combined with single dish observations in order to account for extended emission. The single dish observations 
were carried out at the 45~m telescope at the Nobeyama Radio Observatory (NRO45m). The NRO45m data were taken using two different receivers; BEARS between 2007 and 2009 for $^{12}$CO, and between 2013 and 2014 for \13co and \c18o, and FOREST between 2014 and 2017. These observations are  described in detail in \citet{Shimajiri2014} and \citet{Kong2018}, respectively. For combination and joint deconvolution of the interferometer and single dish data in the \textit{uv}-domain the method presented by \citet{Koda2011} was followed. The data combination results in 0.47~K rms noise level in 0.22~\kms \ channels. Details on the imaging are presented in the CARMA-NRO Orion Survey data paper \citet{Kong2018}.

\begin{figure*}[!ht]
   \centering
   \includegraphics[width=\textwidth]{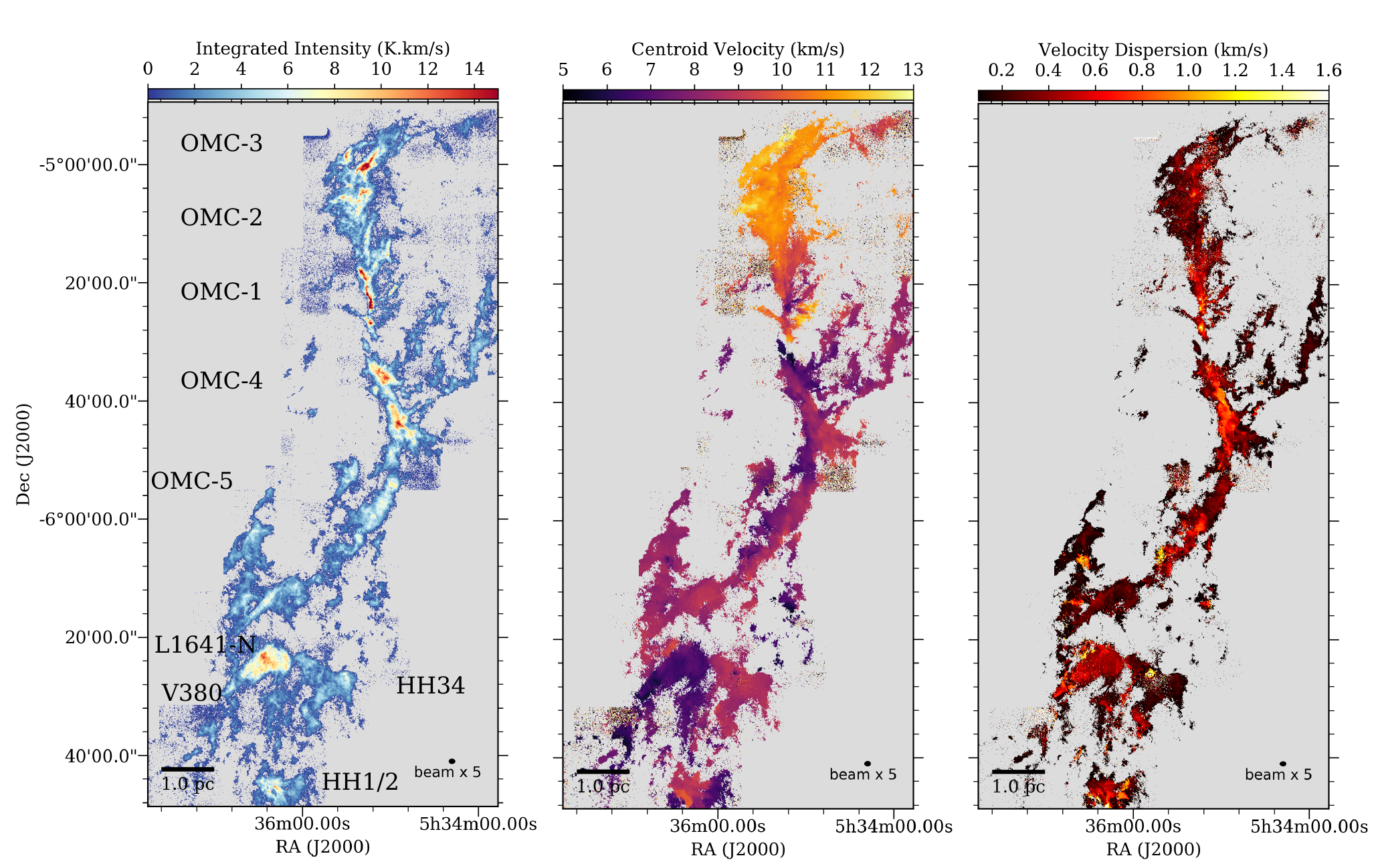}
      \caption{Moment maps of \c18o J = 1--0. From left to right the panels display the integrated intensity emission, the centroid velocity, and the velocity dispersion. Relevant star-forming regions are marked on the integrated intensity map. Beam-size shown in the bottom left corner is five times the real beam-size (5x0.015~pc) of the observations for demonstration purposes.
      %\todo[author=JaimePineda, inline]{the color map used should be swapped... the integrated intensity could use the black to yellow colormap, while the centroid velocity should use the blue to red map.
      %I would show the actual beam (even if it is ridiculously small), rather than the enlarged version.}
      }
             
         \label{fig:fullmap}
   \end{figure*}

The combined (CARMA+NRO45m) data spans 2 degrees in declination including the ISF in the north and the L1641 cloud in the south. In Figure~\ref{fig:fullmap}, we show the \c18o moment maps; integrated intensity, centroid velocity and velocity dispersion (from left to right). 
%We analyze the C$^{18}$O datacube with 2\,arcsec pixels and velocity channels of 0.22\,\kms.
 The data highlights the filamentary nature of the cloud (see Figure~\ref{fig:fullmap}, left panel), as well as hub-like structures where filaments merge. A plethora of kinematic information accompanying the filamentary structure can be seen in the velocity centroid map (Figure~\ref{fig:fullmap}, middle panel). 
In addition to the well known north-south velocity gradient from 13\,\kms to 5\,\kms, we see gradients perpendicular to the spine of the ISF appearing in OMC-4 and -5 regions. The kinematics of the filamentary structure will be discussed in a following paper.  

\section{Filament properties}\label{sec:filamentProperties}
Filament identification is done by employing the structure identification algorithm \textit{Discrete Persistent Structures Extractor}, DisPerSE \citep{Sousbie2011}. DisPerSE uses the Morse theory to derive information on the topology of a given data set. It builds up a tessellation consisting of critical points -- those points where the gradient of the intensity goes to zero, namely: minima, maxima and saddle points. It connects maxima and saddle points via arcs, which are then called \textit{filaments}. DisPerSE has been a widely used tool to extract filamentary structures both in observational and synthetic datasets \citep[e.g.][]{Arzoumanian2011,Schneider2012,Palmeirim2013,Panopoulou2014,Smith2016,Chira2017}. In order to identify filaments in the CARMA-NRO Orion Survey we use the \c18o datacube. The \c18o emission allows us to trace denser regions of the cloud. It is optically thin and less affected by stellar feedback compared to $^{12}$CO and \13co, therefore it is a better choice to trace dense filaments. The $^{12}$CO emission traces more diffuse and optically thick parts of the cloud (see Figure~5 in \citealt{Kong2018}), and therefore we refrain from using it for the analysis of the filamentary structure. The \13co map shows filamentary structure in addition to the extended emission (see Figure~6 in \citealt{Kong2018}), but it can be optically thick in the densest parts of the ISF.

We apply DisPerSE on the \c18o datacube with a persistence and a noise threshold of 2~K (4~$\sigma$). Persistence is a term that refers to the contrast between the values of the identified critical points. 
High persistence means that the identified maximum has higher contrast with respect to its surroundings. Eliminating those critical point pairs with low persistence, in turn, leaves us with the most prominent structures that stand out from the rest. For further discussion on persistence and the general mathematical background of the algorithm we refer to \citet{Sousbie2011}. 

After running on a 3-dimensional position-position-velocity (PPV) cube, DisPerSE provides us with filament coordinates for each identified filament on the PPV space. This has an advantage over %AG
the column density filaments identified in \textit{Herschel} maps \citep[e.g.][]{Andre2016}, because we are able to disentangle the velocity structure of the filaments instead of looking at integrated emission over all the structure along the line of sight. 

\begin{figure}%[t]
   \centering 
   \includegraphics[scale=0.5,trim={80 50 100 10}]{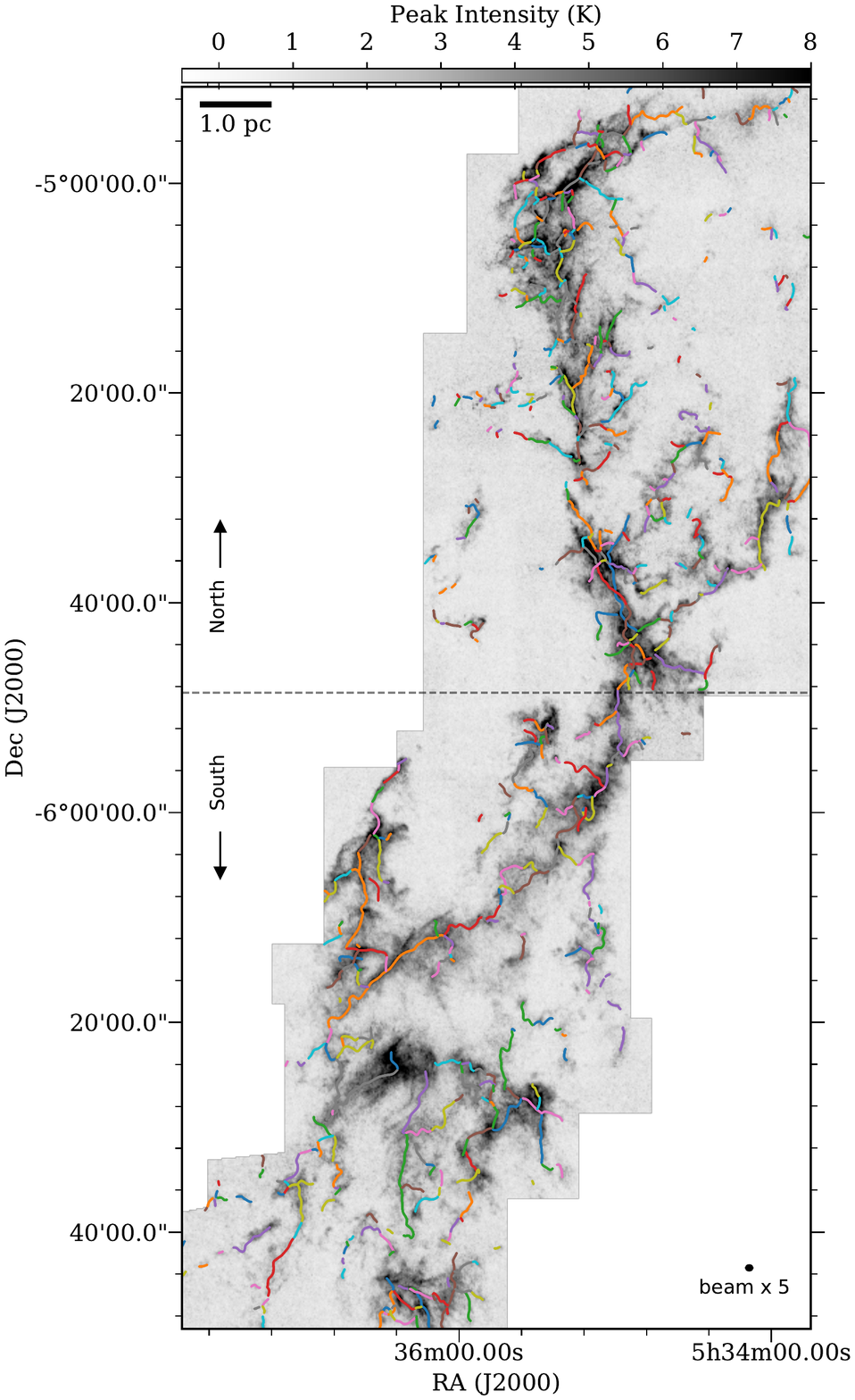}
      \caption{Filaments identified in the Orion A molecular cloud. Greyscale shows the \c18o peak intensity emission, while the colored segments indicate the filaments identified by \texttt{DisPerSE}.  
              }
        \label{fig:filaments}
 \end{figure}
 
Figure~\ref{fig:filaments} shows the identified filaments overlaid on the \c18o peak intensity map. In total, 347 filaments are identified in the northern region (ISF) and 278 filaments in the southern region (northern part of the L1641 cloud). The properties of these filaments are unaffected by the recent distance measurements to the cloud that showed a distance gradient throughout Orion A \citep{Grossschedl2018}. The gradient starts being effective below galactic longitude 210$^{\circ}$ (or approximately $-$6$^{\circ}$30$\mathrm{'}$) close to the southern border of our map. The filaments show slightly different properties in the northern part of the cloud compared to the more diffuse filaments identified in the southern part (see the following sections and Figures~\ref{fig:width_results_all} and \ref{fig:peaksVSwidth}). A large number of filaments identified by DisPerSE have short lengths (< 0.15 pc), and seem to be continuations of one another, resulting in a longer but fragmented (as seen in \c18o) filament. We therefore do not exclude them from further analysis, in particular for the width determination. However, properties of only those filaments with an aspect ratio larger than 2:1 are shown in the Appendix~\ref{appendix:additional}. 

In order to study filament properties, we have developed a python package: the Filament Characterization Package \texttt{FilChaP}\footnote{Publicly available at https://github.com/astrosuri/filchap} that utilizes the coordinates of the identified filaments and derives their physical properties. In the following section, we present the analysis performed by \texttt{FilChaP}.

\subsection{Calculation of the filament width}{\label{sec:width}}
For each filament, we extract radial intensity profiles perpendicular to the filament spines. This is done by first calculating coordinates of a perpendicular line at each sample point along each spine. These sample points are separated by 1.5 beam sizes (0.023~pc) in order to provide statistical independence. This separation value was found by visual inspection of the  pixels employed to produce intensity profiles, making sure the same pixels are not used multiple times in consecutive slices. However, depending on a filament's curvature, the slices may occasionally cross one another at larger radial distances from the spines. The pixels that lie on or adjacent to the calculated perpendicular lines are used to produce the radial intensity profiles in a similar fashion to \citet{Duarte2016}. 

\begin{figure*}[h!]
\centering
\includegraphics[width=\textwidth,height=0.8\textheight,keepaspectratio]{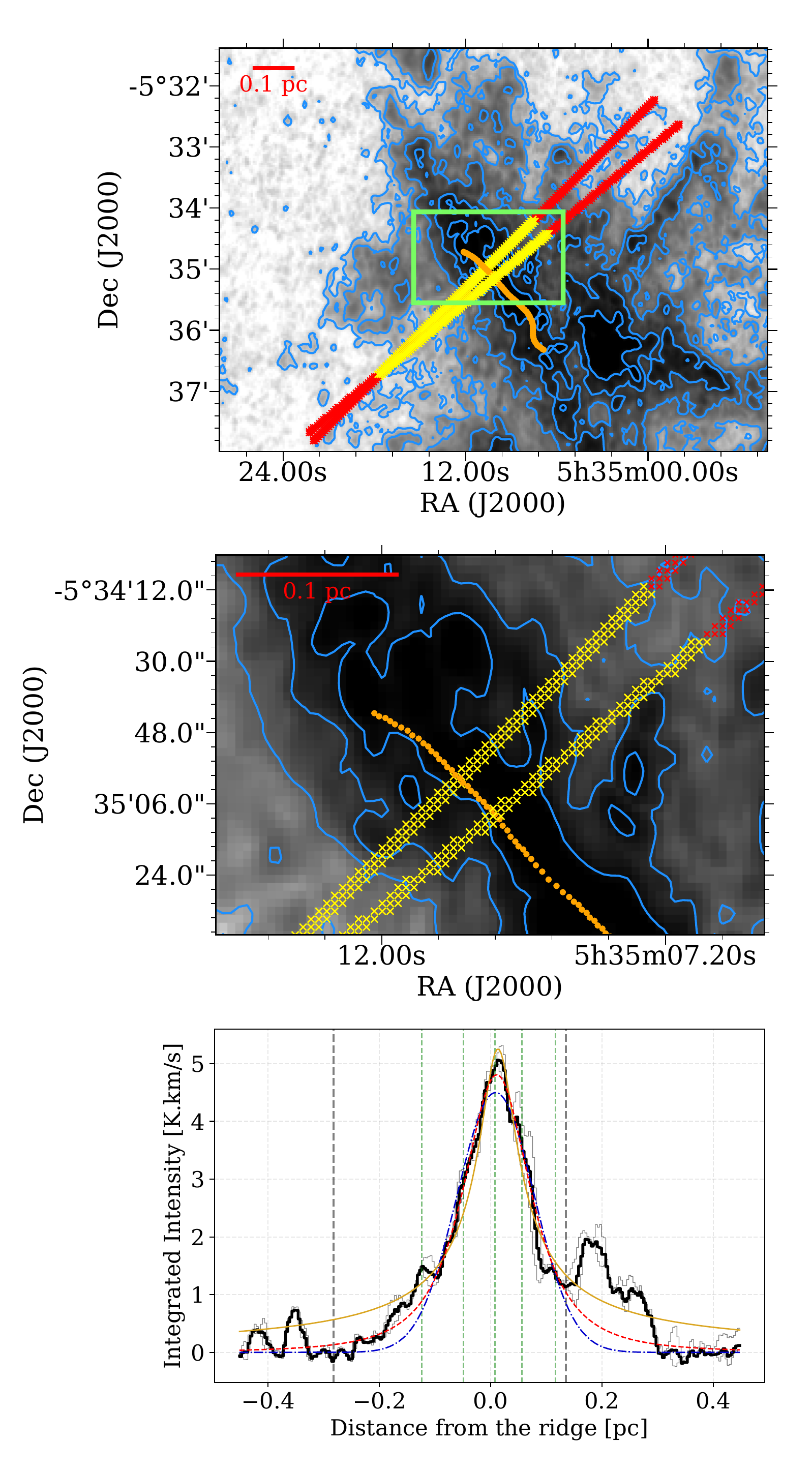}
\caption{\textit{Top}: \c18o peak intensity emission overlaid with a filament spine (in orange) and 2 perpendicular slices (in red and yellow). The yellow portion of the slices indicate the extent of emission fitted with \texttt{FilChaP}. The green box marks the zoom-in region shown in the middle panel. Contours indicate emission levels of 2, 3, 5, 7, 9, 11~K. \textit{Middle}:  Expanded view of the perpendicular slices. The perpendicular pixels used to extract the intensities at each radial bin are marked with crosses. \textit{Bottom}: The intensity profiles (grey lines) extracted from the slices overlaid on the map in the middle panel.  Blue, yellow and red curves fitted to the averaged intensity profile (black) indicate Gaussian and Plummer-like fits with p=2 and 4, respectively. The vertical grey dashed lines show the positions of the identified minima considered during the fitting procedure (see Sect.~\ref{sec:width}). Vertical green dashed lines correspond to shoulders as explained in Sect.~\ref{sec:shoulders}. }
\label{fig:width_example}
\end{figure*}

This process is shown in Figure~\ref{fig:width_example}. On the top panel of the figure the peak intensity map towards the OMC-4 region that is overlaid with an identified filament spine is shown. Two perpendicular slices from which we extract intensity profiles are also overlaid on the filament spine in red.
An intensity profile is constructed using intensities of the pixels that are perpendicular to the filament spine at a sample spine point. These ``perpendicular pixels'' are found on or at the immediate vicinity of the calculated perpendicular line by looking for pixels that are at most 1 pixel away from the perpendicular line. The middle panel of Figure~\ref{fig:width_example} shows an expanded view of the region marked with a green box in the rop panel, in order to provide a more detailed look into the perpendicular slices. The perpendicular pixels are shown with red crosses marking the center of each pixel. There are occurrences where the calculated perpendicular lines do not cross exactly pixel centers. In that case, \texttt{FilChaP} takes the mean of the intensities from pixels that lie adjacent to the perpendicular lines. A complete radial intensity profile is achieved by plotting intensities from perpendicular pixels at each radial distance from the filament spine.

The bottom panel of Figure~\ref{fig:width_example} shows the averaged (over the shown two perpendicular slices) and baseline subtracted\footnote{see Appendix~\ref{appendix:filchap} for baseline subtraction.} intensity profile with a black solid line. In order to obtain the averaged profiles, we take the mean of the intensities at each radial distance from the spine. The negative distances from the spine point to the north-east direction, and positive distances to south-west of the filament. The grey thin lines indicate the different profiles extracted from each slice. Blue, yellow and red curves fitted to the averaged intensity profiles in the right panel denote Gaussian and Plummer--like fits with p=2 and 4, respectively. Grey dashed lines indicate the fitting boundaries. Green dashed lines indicate the shoulders of the averaged profile. A shoulder is identified where there is a local minimum in the second derivative of an intensity profile. Within \texttt{FilChaP} the shoulders are always identified on the smoothed intensity profiles. For this study, the smoothing length is the beam size. We also require the integrated intensity of a shoulder to be at least five times the noise level of the averaged profile which on the order of 1~K.~km~s$^{-1}$. This threshold is set in order to discard less significant peaks.

As each point on the identified filament is defined by a velocity coordinate as well, we know the velocity range spanned by our filament. This information allows us to disentangle filamentary structure in PPV space. Therefore, we utilize this information when calculating the radial intensity profiles in the following way. The intensity profiles are not calculated on the integrated intensity map that is obtained from the entire velocity range of the cloud (as was done in the dust continuum studies). Instead, the intensities from each perpendicular slice are integrated over the two neighbouring velocity channels around the velocity at which the filament is identified. For example, if a point along the filament is at a velocity of 5.5~km~s$^{-1}$ the intensity profile that is perpendicular to this point is calculated from the integrated intensity {\it between} 5.28~km~s$^{-1}$ {\it and} 5.72~km~s$^{-1}$ with the velocity channel width being 0.22~km~s$^{-1}$. This helps reduce the confusion in PPV space by discarding emission other than that which belongs to the identified filament and it makes sure that the emission included in the construction of the intensity profiles is not separated from the velocity at which the filament is identified by more than 2 times the sound speed (approximately 0.2~km~s$^{-1}$ at 10~K , close to our velocity resolution), providing coherence. The velocity coordinate along the filaments often change, therefore, the integrated intensities are computed individually for each PPV point along a filament.

\subsubsection{Filament width determination methods}
In order to calculate the widths, we employ four different methods, a Gaussian fit, two fits with Plummer--like functions with power law indices p=2 and 4, and the second moment of distribution. Plummer-like functions have been used to describe the column density of a filament with a dense and flat inner portion and a power-law decline at larger radii \citep{Arzoumanian2011}:
\begin{equation}
	\Sigma_{p}(r) = A_{p}\frac{\rho_c R_{flat}}{\left[1 + \left(\frac{r}{R_{flat}}\right)^2\right]^{\left(\frac{p-1}{2}\right)}}
\end{equation} 

\noindent where $\rho_c$ is the central density, $R_{flat}$ is the radius within which the density is uniform, $p$ is the power law index, and $A_{p}$ is a finite constant that is related to the filament's inclination compared to the plane of the sky (here we assume that this angle is zero). The $n^{th}$ moment of a distribution $I$ is given by:
\begin{equation}
  \label{eq:width}  
  m_n = \frac{1}{N}\frac{\sum_{i}^{N} I_i \left(x_i-\bar{x}\right)^n}{\sigma^n}
\end{equation}

\noindent where $\bar{x}$ is the intensity weighted mean position of the profile, ${I_i}$ is the intensity at position $x_i$, and $\sigma$ is the intensity weighted standard deviation of the profile. The second moment is the variance of the distribution and therefore we calculate the width of the profile as $\sqrt{m_2} \times 2\sqrt{2ln2}$. For the fits, we use the curve\_fit function contained within the scipy.optimize package with the default Levenberg–Marquardt algorithm.

We refrain from fitting the power law index of the Plummer-like function due to the degeneracy between $p$ and $R_{flat}$. Instead, we use the literature values of the index; p=4 for the density of a homogeneous isothermal cylinder \citep{Ostriker1964} and p=2, the value derived in \textit{Herschel} studies when p is fitted as a free parameter \citep{Arzoumanian2011}, which is attributed to non-isothermality of these objects \citep[e.g.,][]{Lada1999} or to magnetic fields \citep[e.g.,][]{Fiege2000}.  

The profiles are fitted only within the boundaries set by the minima around the peak. These boundaries are shown in the right hand side of Figure~\ref{fig:width_example} as grey vertical dashed lines. The minima are automatically calculated through \texttt{FilChaP} by first smoothing the profiles with a Gaussian having a width of the beam-size of the observations, and then looking at minima around the peak with a significance of $\sim$ 3 beam-sizes. If multiple minima are detected on each side of the spine we take the ones that are located closer to the maxima as boundaries. The effect of the fitting range on the calculated widths is further discussed in Section~\ref{sec:fitting_range}.

\subsubsection{Treating the bias caused by the averaging length}
\begin{figure*}
   \centering
   \includegraphics[width=\hsize]{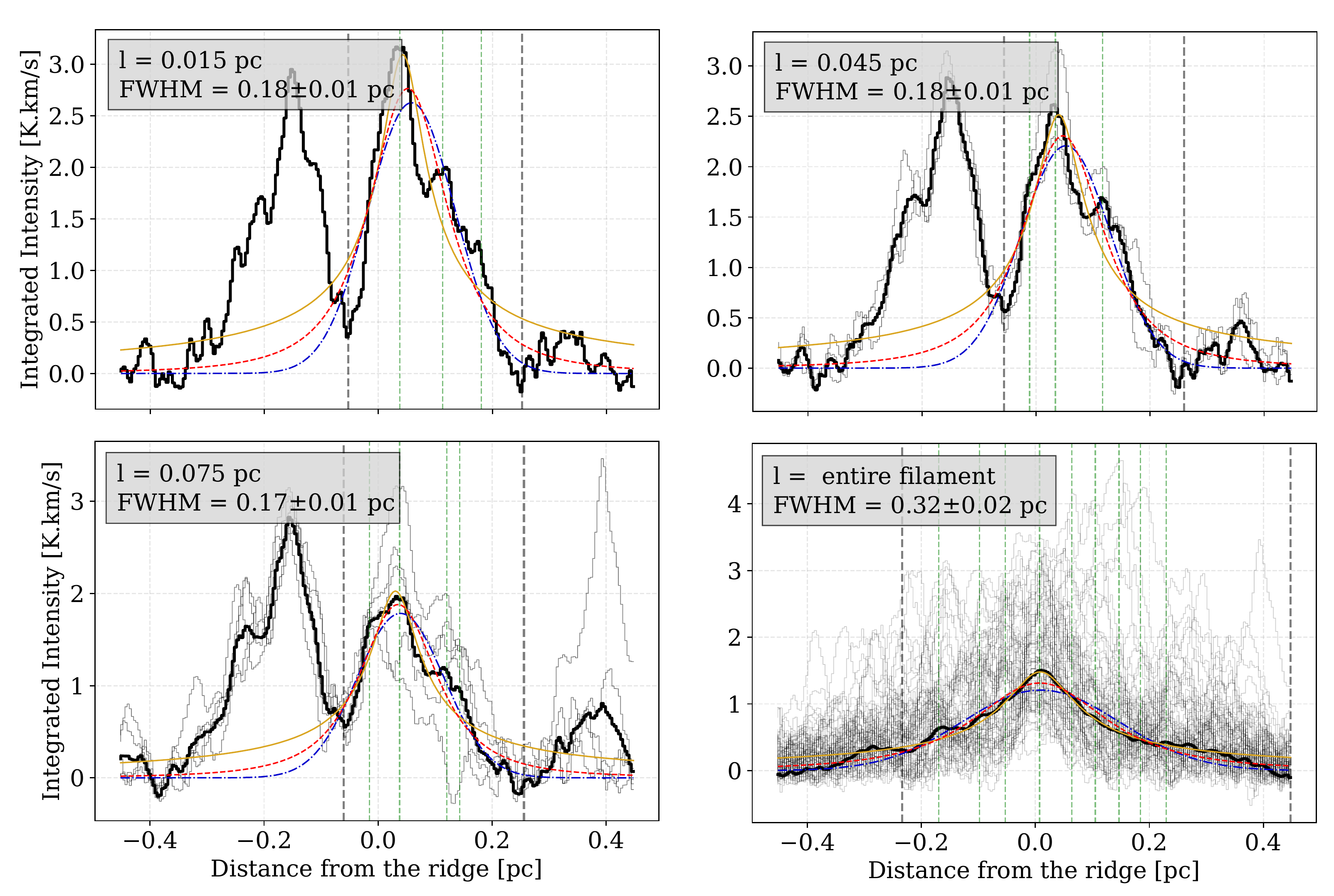}
      \caption{Plots showing the effect of averaging intensity profiles along different length scales. Individual intensity profiles are shown in grey, while averaged intensity profiles are shown in black. Blue, yellow and red curves fitted to the averaged intensity profile (black) indicate Gaussian and Plummer-like fits with p=2 and 4, respectively. The grey vertical dashed lines indicate the fitting boundaries and the green vertical dashed lines indicate the shoulders identified by \texttt{FilChaP}. From the top left to bottom right the profiles are averaged over 0.015, 0.045, 0.075~pc and the entire filament with corresponding Gaussian FWHM of 0.18, 0.18, 0.17 and 0.32~pc.}
         \label{fig:width_average}
\end{figure*}

Previous studies \citep[e.g.][]{Arzoumanian2011} calculated filament widths using intensity profiles that are averaged over the entire length of a filament. Although averaging helps to obtain a smooth radial intensity profile, it has the disadvantage that any information that could be used to understand how and why the width changes along a single filament is lost. Therefore, we first study the effect of averaging intensity profiles by averaging over a length of 0.015~pc (a beam-size), 0.045~pc, 0.075~pc and finally over the entire filament. Figure~\ref{fig:width_average} shows how averaging intensity profiles over different length scales affects the results. The length scale over which the profiles are averaged is indicated on the top left corner of each panel. The panels where we average over shorter length scales (top left and right, bottom left) show marginal change in width. The bottom right panel, however, reveals if we were to average over the length of the filament, the resulting width is twice as large as the previous values. Thus, not only do we overestimate the filament width by averaging, but we also lose information on the sub-structure of the filament. We then choose to average intensity profiles over a length of 0.045~pc (3 beam-sizes) for further analysis. While this approach gives us a chance to smooth the profiles over 12 points (3 beam-sizes) along the filament, it helps us avoid deriving overestimated parameters. Furthermore, we are able to follow the gradient of the width along the identified filaments. 

%%%%%%%%%%%%%%%%%%%%%%%%%%%%%%%%%
%%% figure on the width averaging
%%%%%%%%%%%%%%%%%%%%%%%%%%%%%%%%%

\begin{figure*}
   \centering
   \includegraphics[width=\hsize,trim={10 20 10 10}]{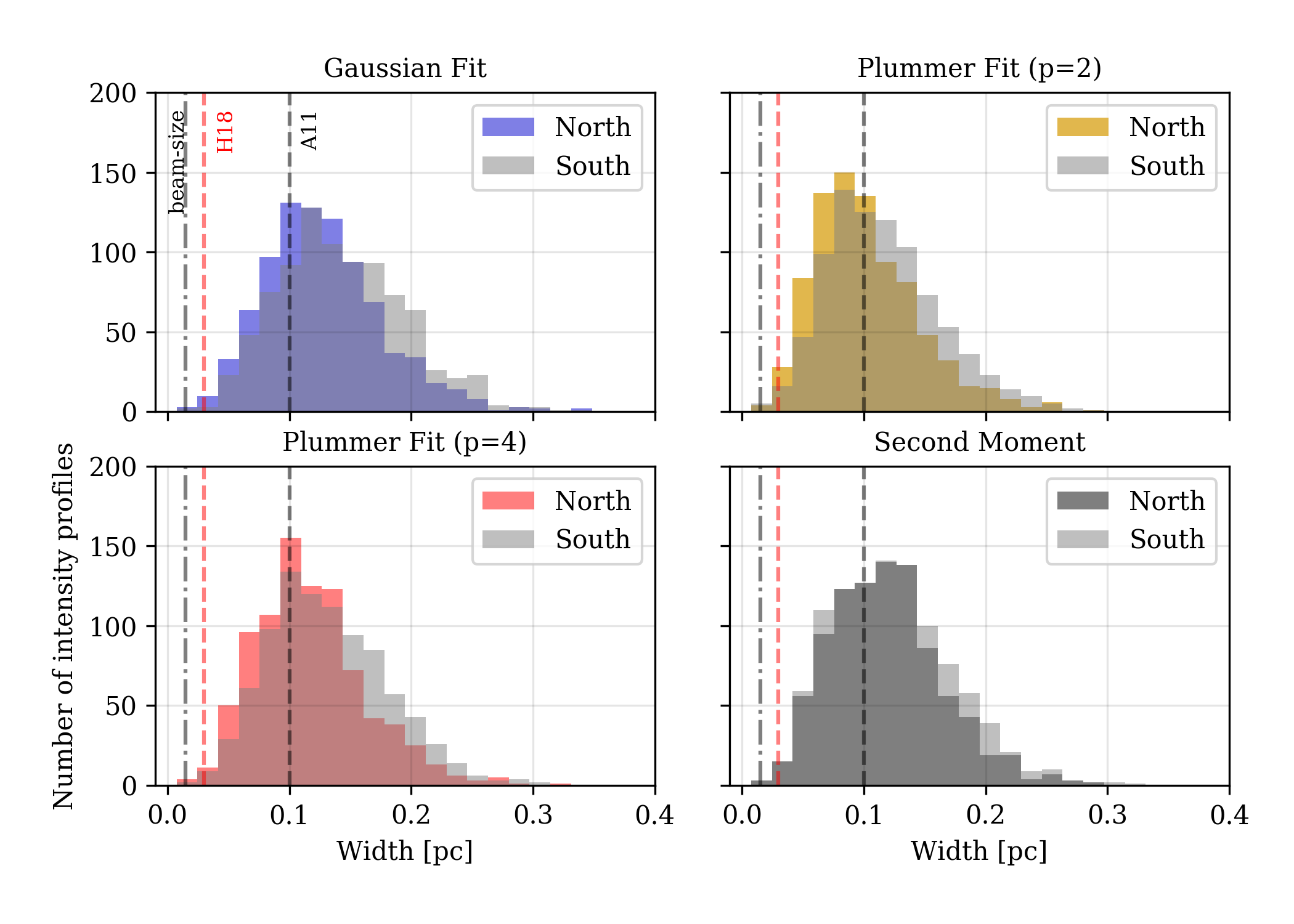}
      \caption{Distribution of filament widths in the norther (colored) and southern (grey) regions of Orion~A (see Fig.~\ref{fig:filaments}) calculated using Gaussian fits (top left), the second moment of distribution (bottom right), and Plummer-like fits with power law indices p=2 (top right) and p=4 (bottom left). The vertical dashed lines represent the filament widths from \citet{Hacar2018} in red, and the ``characteristic'' filament width  \citep{Arzoumanian2011} in black. The dot-dashed line indicates the beam-size of our observations.}
         \label{fig:width_results_all}
\end{figure*}

\subsection{Distribution of filament widths}
Using the method described in Section~\ref{sec:width}, we calculate the widths of all the 625 filaments identified by DisPerSE in Orion A. The results are shown in Figure~\ref{fig:width_results_all} where the distribution for the northern region is  shown by the colored and the southern region by the grey histograms. The four subplots show the distribution of widths (FWHM) calculated using Gaussian (left) and Plummer p=2 fits (right) in the top panel, Plummer p=4 fit (left) and the FWHM calculated from the second moment of distribution (right) in the bottom panel. Black and red dashed lines in each subplot plot indicates the ``characteristic'' widths found by \citet{Arzoumanian2011} in Herschel dust continuum, and \citet{Hacar2018} in N$_2$H$^+$ emission. The dot-dashed line indicates the spatial resolution of the data. 

All four methods yield similar distributions, with the Plummer p=2 fits resulting in slightly narrower widths\footnote{Using the deconvolution formula given in \citet{Konyves2015}, the deconvolved filament width would be $\sqrt{0.1^2-0.015^2}$= 0.098~pc using an observed width of 0.1~pc and our beamsize of 0.015~pc. This shows that for our high-resolution data the beamsize does not play a crucial role for our calculated filament widths and there is no need for deconvolution}. 
The median values\footnote{All the errors cited on the median widths throughout this work are the first and third quartiles. The uncertainties on the widths derived from the intercept of the linear relation between number of shoulders and the width (see Section~\ref{sec:shoulders}) come from the error on the fitting.} of the distributions are $0.12\substack{+0.15 \\ -0.09}$~pc and $0.14\substack{+0.18 \\ -0.11}$ ~pc for the Gaussian fits, $0.09\substack{+0.13 \\ -0.07}$ ~pc and $0.11\substack{+0.14 \\ -0.08}$ ~pc for the Plummer p=2 fits, $0.11\substack{+0.14 \\ -0.08}$ ~pc and $0.13\substack{+0.16 \\ -0.09}$ ~pc for the Plummer p=4 fits, and finally, $0.12\substack{+0.15 \\ -0.09}$ ~pc and $0.12\substack{+0.15 \\ -0.09}$ ~pc for the FWHM derived from the second moment for the northern and southern regions, respectively. 
These results are very close to what has been reported by the \textit{Herschel} studies \citep[e.g.][]{Arzoumanian2011, Andre2017} and are larger than the width reported based on ALMA studies of filaments \citep[e.g.][]{Hacar2018, Henshaw2017}, which is not unexpected considering that N$_2$H$^+$ is less extended and traces higher densities along filament spines.  

\begin{figure*}[t]
   \centering
   \includegraphics[width=\hsize]{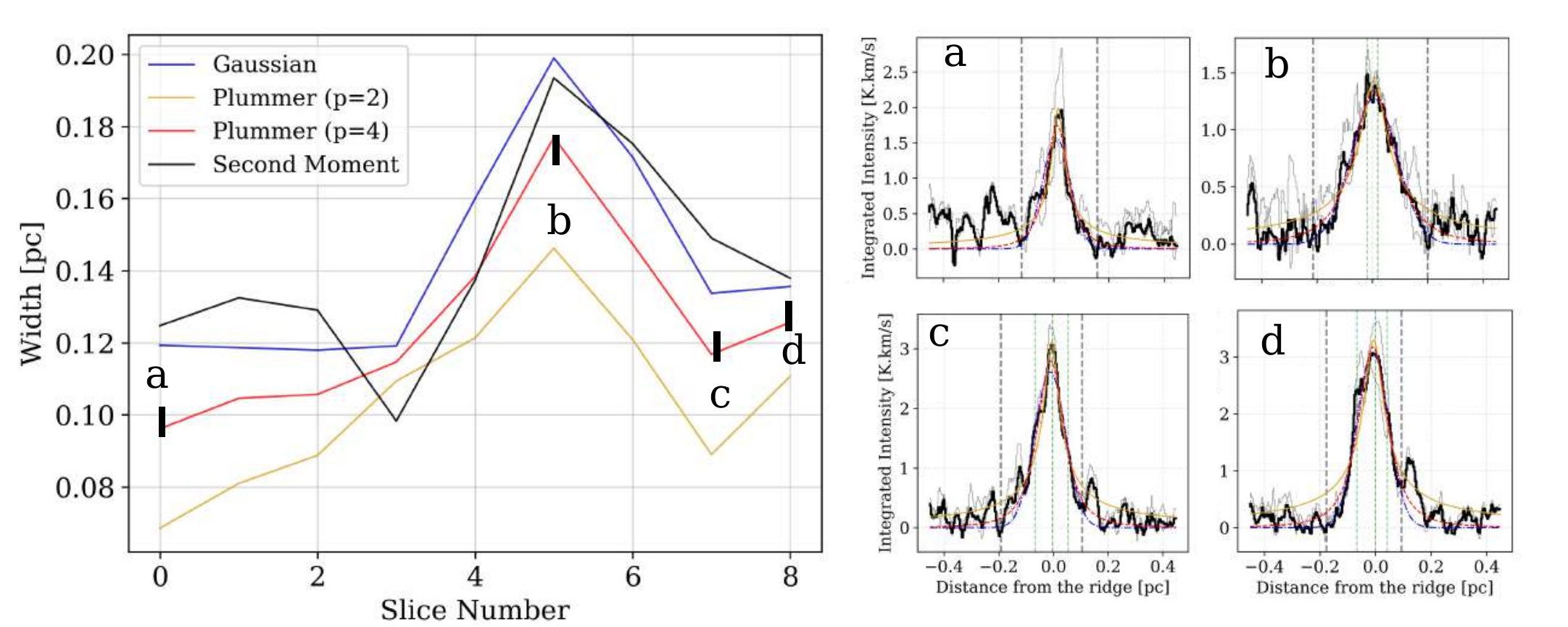}
      \caption{\textit{Left}: Variation of filament width along the length of a single filament taken as an example. The slice numbers correspond to different positions along the filament. Each slice shown here represents the averaged value over two consecutive perpendicular cuts. These cuts are always separated by 1.5 beamsizes (0.0225~pc) and the length of the filament is 0.4~pc. \textit{Right}: Selected intensity profiles at points ``a'', ``b'', ``c'' and ``d'' show in the left panel. These points are selected to best represent the variation of the width along the filament. Vertical grey dashed lines indicate the boundaries of the fits while the green dashed lines indicate the shoulders of the profiles. }
         \label{fig:example_width_gradient}
\end{figure*}

The existence of a ``characteristic'' filament width is still highly debated (see \citealt{Pano2017, Andre2017}). In studies such as \citet{Arzoumanian2011,Palmeirim2013,Koch2015} the spread of the filament width distribution is less than a factor of two around the median width. On the other hand, a study by \citet{Juvela2012} shows that there is an order of magnitude deviation from the mean width (0.1--1.0~pc). Our distributions shown in Figure~\ref{fig:width_results_all} indicate a very significant, about an order of magnitude spread around the median value regardless of the method. The origin of this wide distribution may be due to the fact that the filaments are not isolated homogeneous cylinders. They may have variations along their length that are not taken into account if the entire filament is represented by an averaged number. In Figure~\ref{fig:example_width_gradient} we show an example of how the width changes along \textit{a single filament}. In the left panel, we plot the slice number along the filament (cross-sections for radial intensity profiles) against the calculated width at the corresponding slice. The Plummer p=2 fit yields narrower widths along the filament, this agrees with the general picture that we have drawn from Figure~\ref{fig:width_results_all}. The right panel shows 4 subplots corresponding to 4 intensity profiles along the filament that are marked by ``a'', ``b'', ``c'' and ``d'' in the left panel. These intensity profiles are selected to be representative of how the width changes. At point ``a'', the intensity profile is the narrowest and peaky with a width of 0.12~pc (Gaussian FWHM). As we progress from point ``a'' to ``b'' we see that the substructures (apparent at around 1 K.~km~s$^{-1}~$, close to five times the noise level of the averaged profile) that appear at point ``a'' get wider, hence the width increases. At point ``b'' the intensity profile is broad and the width is the largest, reaching 0.2 pc. From point ``c'' to ``d'' the shoulders on both sides of the peak (shown in green dashed lines) get wider and acquire higher intensities, which in turn causes the increase in width. The unresolved inhomogeneities within the filament itself that can be traced by these shoulders in the profile cause the filament's width to vary. 

\subsubsection{Goodness of the fits}\label{sec:goodness}
\begin{figure*}[t]
   \centering
   \includegraphics[width=\hsize]{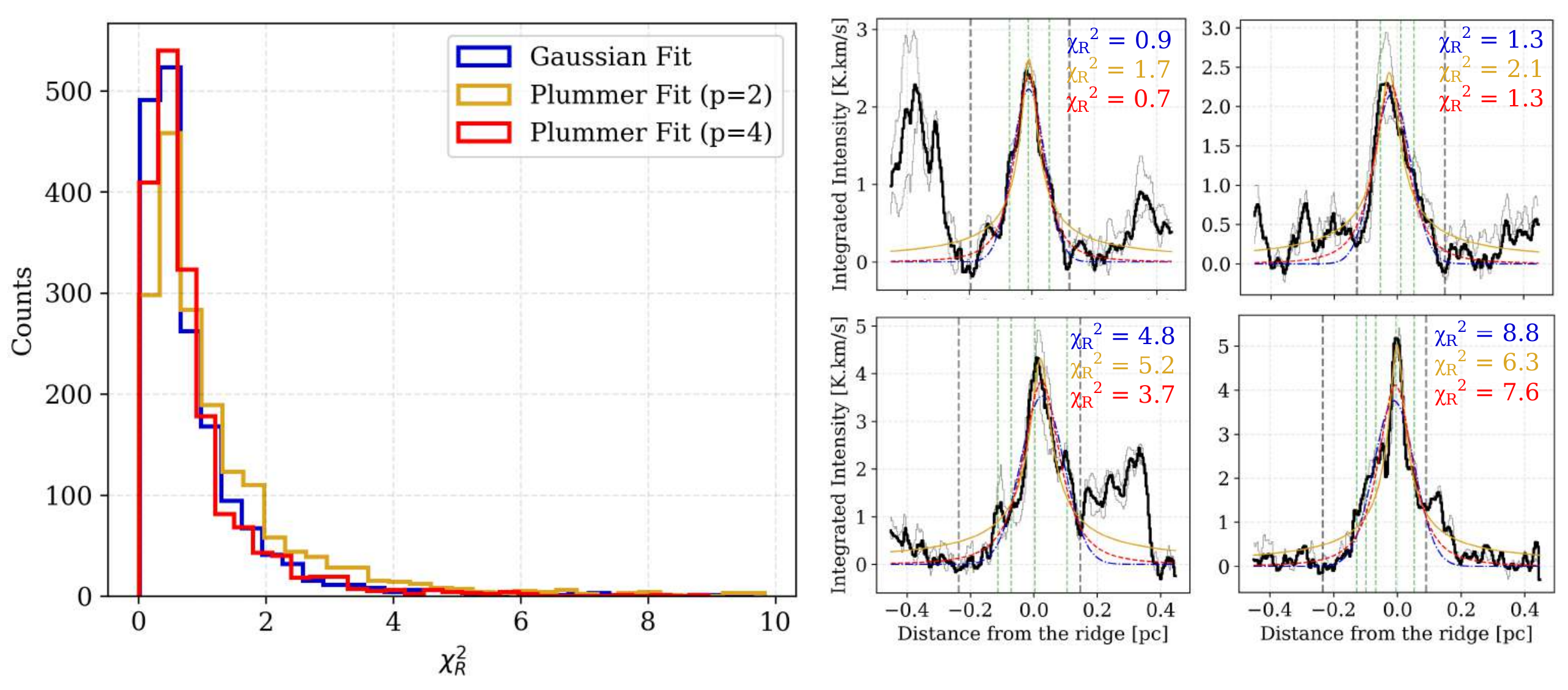}
      \caption{\textit{Left}: ${\chi}_R^2$ values for each radial intensity profile fit for the different methods used in the width determination. \textit{Right}: Four examples of profiles with different reduced chi-squared values. The bottom right sub-panel shows the example of a fit discarded due to its large reduced chi-squared.}
         \label{fig:chi_sq}
\end{figure*}

In order to assess the goodness of our radial intensity profile fits we calculate the reduced chi-squared values for each individual fit and for the three different methods that we employed. The reduced chi-squared is simply calculated as chi-squared per degree of freedom\footnote{We have 3 degrees of freedom per fit; the amplitude, mean, and standard deviation (or R$_{flat}$ for the Plummer fits).}. The resulting distributions of reduced chi-squared of all the fits are shown in the left panel of Figure~\ref{fig:chi_sq}. The median reduced chi-squared values of our fits are 0.6 for the Gaussian and Plummer p=4 fits, and 0.8 for the Plummer p=2 fits. The reduced chi-squared value for a \textit{good} fit is expected to be close to 1. However, by inspecting the intensity profiles and their corresponding reduced chi-squared values, we see that with the reduced chi-squared values up to 6, the fits still describe the underlying intensity profiles well.  The right panel of Figure~\ref{fig:chi_sq} shows four different intensity profiles, overlaid with the calculated fits and the corresponding reduced chi-squared values for each fit. In both panels of the figure, blue, yellow, and red colors represent the Gaussian, Plummer with p=2 and Plummer with p=4 fits. Profiles with blending structures or extended shoulders such as the example shown in the bottom right subplot leads to poor fits with high reduced chi-squared (${\chi}_R^2>$ 6).  For the Gaussian fits, 98\% of the reduced chi-squared values lie between 0 and 6, while this value is 97\% for the Plummer p=2 and 99 \% for Plummer p=4 fits. We confirm by eye %inspection 
that the tail of the reduced chi-squared distribution (${\chi}_R^2>$ 6) corresponds to
questionable fits. The reason for this is the complexity of the intensity profiles. As the width of these profiles are hard to estimate, they are discarded from further analysis. We note that the reduced chi-squared criterion applies for the Gaussian and Plummer fits individually; i.e. a profile can have a good reduced chi-squared for Plummer p=2 fit but not for the Gaussian, and in this case only the Plummer p=2 width is accounted for. As we do not have a reduced chi-squared value for the widths calculated from the second moment method, we take those second moment widths where the reduced chi-squared criterion is satisfied in all the other three methods. 

\subsubsection{Dependence on the fitting range}\label{sec:fitting_range}
\citet{Arzoumanian2011} and later studies  of filaments \citep[e.g.][]{Andre2016,Federrath2016} refrain from fitting the wings of the column density profiles, and the fitted inner range of the profiles is always the same for each and every column density slice. \citet{Smith2014} and \citet{Pano2017} show that a correlation between the fitting range and the calculated width of these profiles exists. This correlation is not unexpected. Depending on the portion of the profiles that is fitted, the larger the fitted area, the wider the width of the profile will be. 
This behavior arises from the fact that the filament profiles are not isolated. Therefore, the larger the fitting range, the more emission the function will try to fit. In an ideal case, where the baseline was perfectly smooth, the fitting range should not affect the width of the distribution. In real observed data, however, the filament profile is accompanied by extended emission as well as emission from nearby filaments and hubs. Hence, a linear correlation between the fitting range and the resulting width is inevitable. To be as independent of this bias as much as possible, instead of forcing a fixed boundary for our fits, we calculate a unique fitting range for every individual profile. This is done automatically within \texttt{FilChaP} by looking for minima around the peak intensity of the filament. If the minima have a significance of 3 beam sizes (which corresponds to a well resolved nearby or blended structure) their locations mark the boundaries for the fitting range. In this approach, each profile has its own fitting range which depends on the filament's environment. 

\begin{figure*}[t]
   \centering
   \includegraphics[width=\hsize]{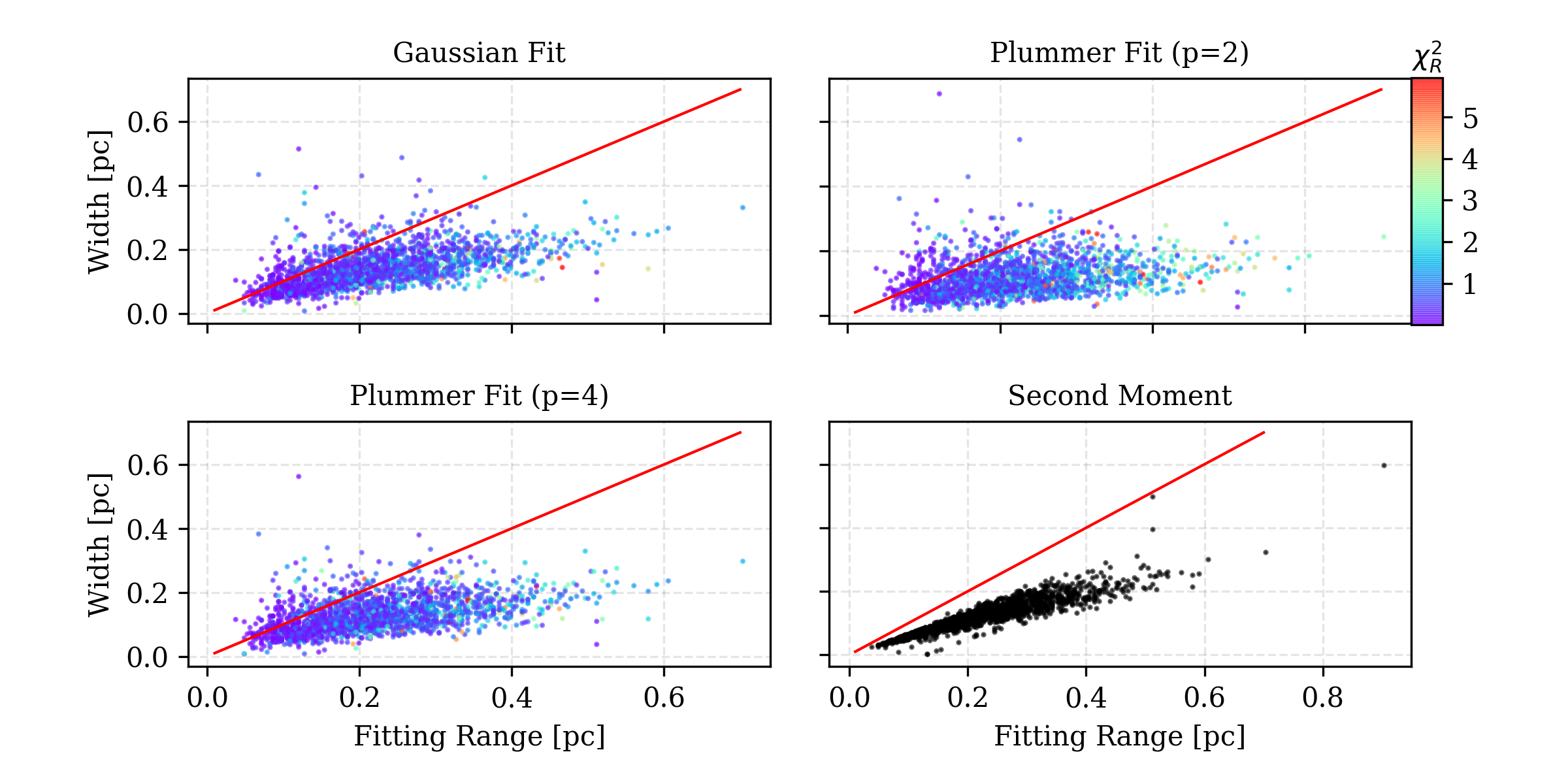}
      \caption{Correlation between the calculated filament widths and the fitting range on the radial intensity profile over which the width is calculated. The red diagonal line shows the 1-1 correlation between the two parameters. The strongest correlation is observed when instead of the fits, the second moment of the distribution is used. The color bar indicates the reduced chi-squared values for the fits. There is also a slight correlation between the reduced chi-squared and the fitting range.}
         \label{fig:range_width}
\end{figure*}

Figure~\ref{fig:range_width} shows the the correlation between the fitting ranges for each individual profile and the derived widths based on the 4 different methods. The scattered points are colored by the ${\chi}^2$ 
of each fit, except for the second moment widths, as those are not fitted but calculated directly from the distribution. The red diagonal lines correspond to a 1-1 relation between the two axes. The strongest correlation between the fitting range and the width is for the case of the second moment (bottom right panel). This is purely because the second moment is calculated based on the boundaries of the distribution, so that the larger the distance between the boundaries the larger is the width. As is also seen from this panel, the second moment width of the distribution cannot be larger than the fitting range. For the Gaussian and Plummer fits, the widths can occasionally be larger than the fitting range, these are the points that lay above the 1-1 correlation line. These points are excluded from the analysis as they originate from 
%incohesive multi-peak 
profiles  with multiple peaks that are not straightforward to analyze.

Of the four different methods used to calculate the filament widths, the Plummer (p=2) fit seems to be the most independent of the fitting range as it shows the shallowest increase with the increasing fitting range. However, as can be seen in the various intensity profiles shown in Figures~\ref{fig:width_example} and \ref{fig:chi_sq}, the Plummer (p=2) only fits the very inner portion of the profile. This is acceptable when one has multiple peaks, but it 
%lacks 
excludes the information on the wings of the profile. This is the reason why the widths calculated from the Plummer (p=2) are always narrower and less dependent on the fitting range than the ones estimated from the other methods. Furthermore, because it fits a substructured profile better, it has fewer points above the 1-1 relation between the fitting range and width. However, we refrain from drawing conclusions of filaments' state of isothermality based on the power-law indices of the Plummer fits, because the difference between the p=2 and p=4 seem to be sensitive to, at least in this work, the amount of background emission observed towards individual filaments.

\section{Discussion}{\label{sec:discussion}}

\subsection{Variation of the filament width across the cloud}{\label{sec:shoulders}}

\begin{figure*}%[t]
   \centering
  \resizebox{\hsize}{!}{\includegraphics[width=\hsize]{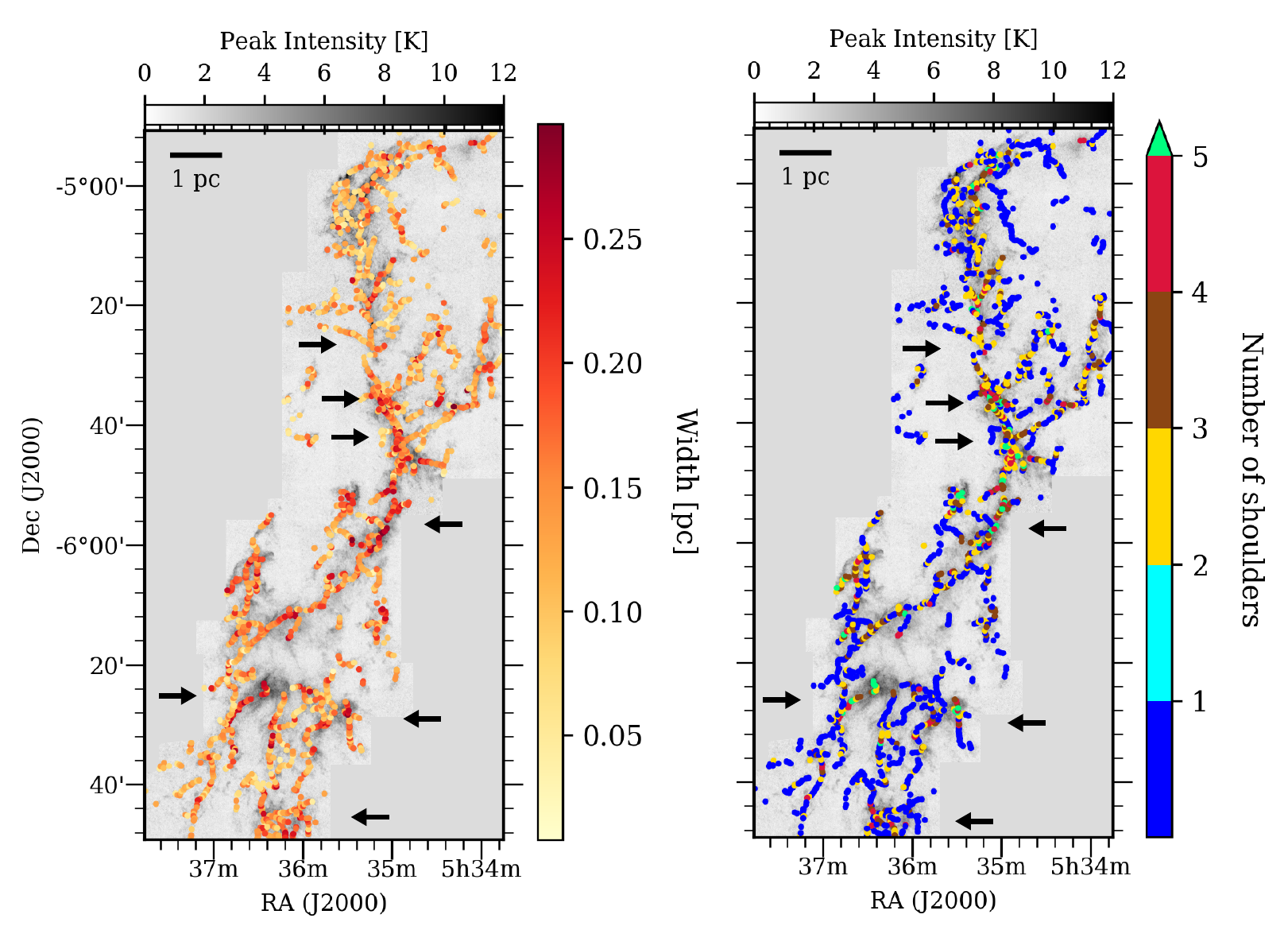}}
      \caption{Filament width (left panel) and number of shoulders (right panel) for the filaments idnetified in Orion~A overlayed on the \c18o peak intensity map.  The color of each point along filaments represents the width (left) and the number of shoulders (right) at that point, corresponding to the colorbar shown in the right side of the panels. The arrows indicate the positions of hubs where filaments converge.}
         \label{fig:gradients_all}
\end{figure*}

As shown in Figure~\ref{fig:example_width_gradient} we see that the filament width varies even along a single filament. This variation is often caused by blending structures. Close to hubs -- where filaments  meet, the intensity profiles get wider as many structures merge into one another. We follow the variation of width along each filament by plotting the width (Gaussian FWHM) on the \c18o map\footnote{The relative width between the slices of a filament is meaningful no matter which method is used (see Appendix~\ref{appendix:filchap}).}. This spatial variation is shown on the left hand side of Figure~\ref{fig:gradients_all} with the points along each filament colored by the width calculated at that point. The corresponding colorbar indicates the calculated width. In order to demonstrate the complexity of the radial intensity profiles we also calculate the number of \textit{shoulders} identified in each profile. The shoulders appear as bumps within the boundaries within which we calculate the width of the filaments, further proving that these radial profiles are not simple single-component Gaussians. Examples of the intensity profiles with detected shoulders are shown in the right hand panels of Figure~\ref{fig:chi_sq}. In this example, the boundaries are shown with the grey dashed lines that clearly separate the emission of the filament from a nearby structure that appears at around 0.2~pc (bottom right subplot). The green dashed lines indicate the locations of the detected shoulders that correspond to bumps within the filament profile that are not significant enough to be separate structures. Up to this point, we have qualitatively shown that filament profiles are substructured and complex which in turn leads to variation of filament widths. A possible correlation between the number of shoulders, the locations of filaments, and the filament widths would indeed allow us to study this phenomenon quantitatively. %Therefore, we 
We plot the identified filaments on the \c18o peak intensity map, this time with each point colored by the number of shoulders detected in the corresponding intensity profile.  Figure~\ref{fig:gradients_all} shows how this number changes along the filaments and across the cloud. 

\begin{figure}%[t]
   \centering
   \includegraphics[width=\hsize]{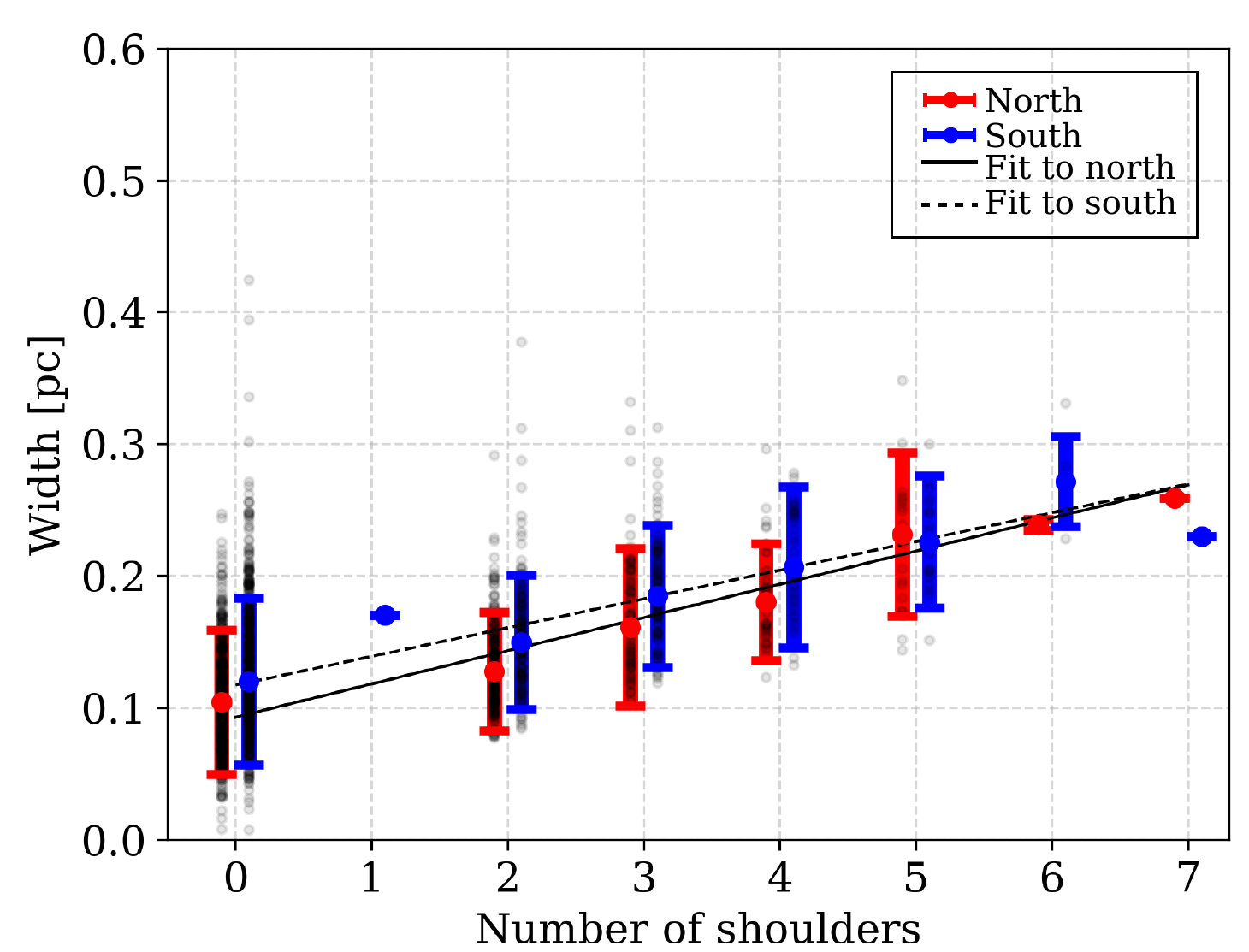}
      \caption{Variation of the filament width with respect to detected number of shoulders in the corresponding intensity profile. The gray circles represent the data points, the red and blue filled circles are the median values. The shown error bars on the median values represent the interquartile range.}
         \label{fig:peaksVSwidth}
\end{figure}

Figure~\ref{fig:gradients_all} 
 suggests a clear correlation between the filament width and the number of shoulders. To demonstrate this correlation better, in Figure~\ref{fig:peaksVSwidth} we show the detected number of shoulders plotted against the corresponding widths of each individual profile. The data points are plotted as grey circles, and their median values are shown as red and blue filled circles for the northern and the southern regions of the map, respectively. The error bars on the median values indicate the interquartile range. The majority of the filament profiles have at least 1 shoulder detected. The existence of such highly complex intensity profiles  implies that filaments are not perfectly isolated cylinders, but that they have substructures within them. Figure~\ref{fig:peaksVSwidth} clearly shows that there is a linear correlation between the number of shoulders and the width: the higher the number of shoulders, the wider is the calculated width. \citet{Clarke17} suggest that filaments may retain wide widths even when supercritical, due to substructure. This substructure lessens the global effects of gravity and produces wider filaments. The linear correlation we find in Figure \ref{fig:peaksVSwidth} supports this view. This correlation suggests widths of 0.09$\pm$0.02~pc and 0.12$\pm$0.01~pc for filaments without shoulders, for the northern and southern region, respectively.

\subsection{The lack of correlation between filament width and column density}
\begin{figure}%[t]
   \centering
   \includegraphics[width=\hsize]{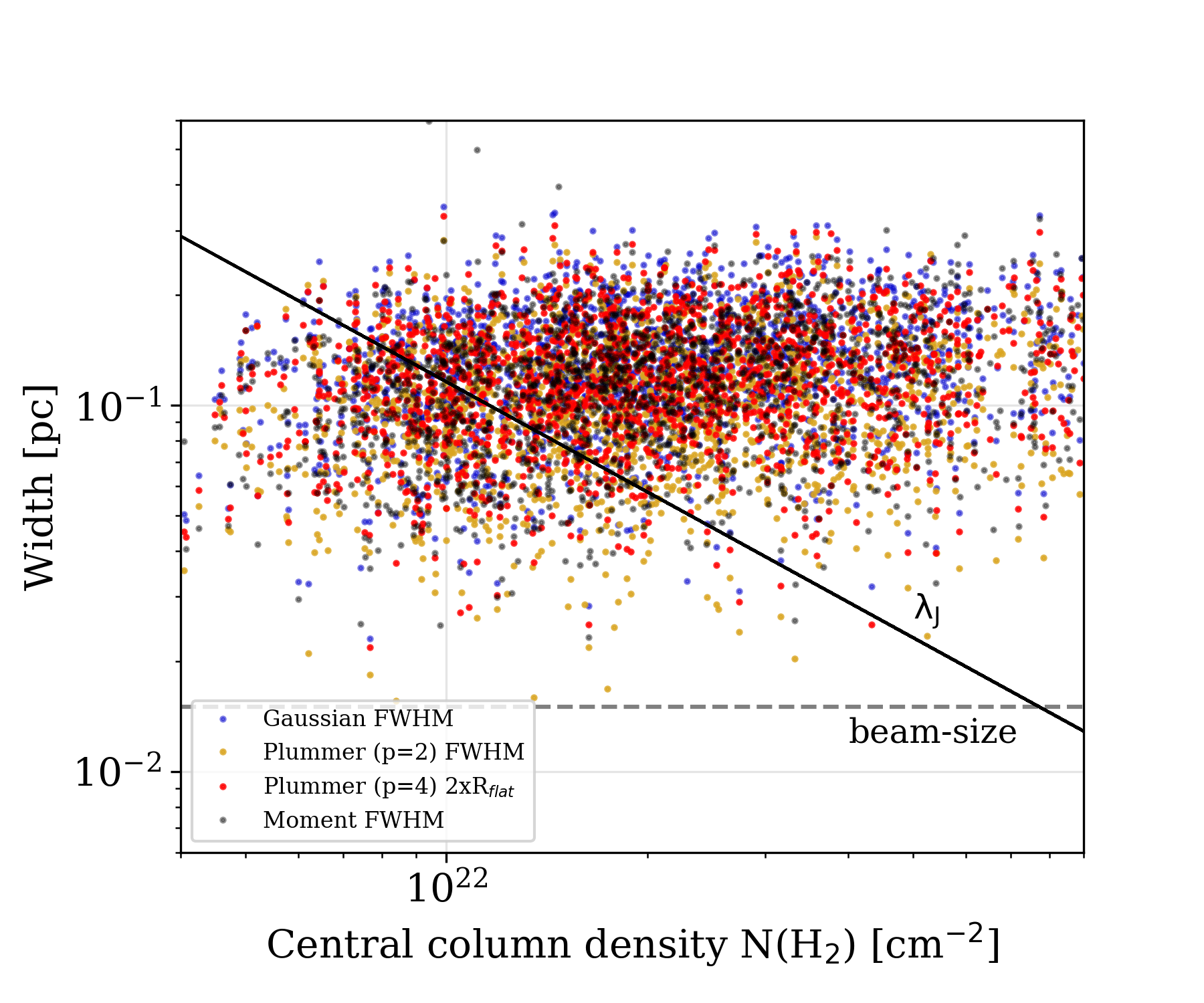}
      \caption{Filament widths calculated using the four different methods plotted against the central column density at each point where the width is calculated. The colors represent FWHM of Gaussian fits (blue), moment analysis (black), Plummer p=2 (yellow) and p=4 (red). The dashed line represents the resolution of our observations and the solid diagonal line the thermal Jeans length at 10~K.}
         \label{fig:widthVScolumnDensity}
\end{figure}

To account for the variation of the filament width throughout the ISF, we also checked for a possible (anti-)correlation between the column density along the filaments and their width. Figure~\ref{fig:widthVScolumnDensity} shows the central column density at every point along the spine of the filaments plotted against the corresponding width at that point. The column density of each pixel along the identified filaments is taken from the \textit{Herschel} dust column density map published in \citet{Stutz2015}.  Blue, yellow, red and black colored points in the plot represent widths derived Gaussian fits, Plummer p=2 and p=4 and the second moment, respectively. The dashed line is the resolution of the observations, and the solid diagonal line is the thermal Jeans length at 10~K; $\lambda_J = c_s^2/G\mu_H\Sigma_0 $ where $\Sigma_0$ is the central column density \citep{Arzoumanian2011}. Ideally, if the filaments are super-critical, and thus gravity-dominated (right hand side of the $\lambda_J $ line), they would contract to fragment and form cores, hence leading to an anti correlation between the width and the central column density. However, this does not appear to be the case. This lack of anti correlation has been found by \citet{Arzoumanian2011} in low-mass star forming regions; Aquila, IC5146 and Polaris. We confirm that even in a high-mass star forming region that is as active and complex as Orion A, the decoupling of filament widths from the central column densities of the filament spine persists. We propose the width variation along filaments is indeed mostly regulated by environmental effects, i.e.~their  positions relative to the hubs and whether the filaments are rather isolated or are connected to or disconnected from surrounding structures. 

\citet{Arzoumanian2011} speculated that the decorrelation of the width and central column density can be explained with turbulent filament formation through compression \citep[e.g.][]{Padoan2001} for the sub-critical filaments, and the super-critical filaments may be able to maintain an approximately constant width if they are continuously accreting from their surroundings. Following this work, \citet{Heitsch2013b} studied gravitational infall onto isothermal filaments that are subject to external pressure to see for which conditions the filament width can be independent of the column density. Their best model that fits the lack of correlation seen in observations is for the magnetized filaments that are subject to varying external pressure due to ram pressure exerted by the infalling material. Our results in Figure~\ref{fig:widthVScolumnDensity} can be directly compared to the results in \citet{Heitsch2013b}, particularly their Figure 4. Even though their best fitting model represents  the majority of our data points, we do not see a decrease in filament width towards higher column densities up to 10$^{23}$ cm$^{-2}$ as the model predicts. 

\section{Conclusions}{\label{sec:conclusions}}
We present a study of the filament properties in the Orion A molecular cloud as seen in the emission of the  dense gas tracer \c18o, produced by the CARMA-NRO Orion Survey \citep{Kong2018}. We identified 625 filaments in the 3D datacube (PPV) using the DisPerSE algorithm. In this first paper, which will be followed by a series of papers including the analysis of filament kinematics and filament properties in different tracers (\13co, dust and C$^+$), we investigated the physical properties of filaments with an emphasis on the filament width. 

Based on the shortcomings of the methods that led to the identification of a ``universal'' filament width, criticized in previous works by \citet{Smith2014} and \citet{Pano2017} we developed an improved and automated method to study characteristics of filaments; \texttt{FilChaP}, a python based algorithm. In this method, while calculating the filament widths: (i)  the radial intensity profiles are not averaged over the length of an entire filament, but over a number of consecutive slices that represent a region 3 beam-sizes in length, (ii) there is no fixed fitting range for these profiles that would force a prefixed width. Instead, each profile has its own fitting range. This range is set by the minima around the peak emission from the filament. The minima are required to have a significance of 3 beam-sizes, (iii) we use 4 different methods to derive filament widths; with a Gaussian fit, two Plummer-like fits with p=2 and  p=4, and the FWHM derived from the second moment of the intensity distribution and (iv) we judge the goodness of each fit by looking at their reduced chi-squared values to prevent `poor' fits biasing the results. We find the following key points:

\begin{itemize}
    \item The median filament widths are $0.14\substack{+0.03 \\ -0.02}$~pc and $0.16\substack{+0.04 \\ -0.03}$ ~pc for the Gaussian fits, $0.11\substack{+0.03 \\ -0.02}$ ~pc and $0.13\substack{+0.04 \\ -0.03}$ ~pc for the Plummer p=2 fits, $0.13\substack{+0.03 \\ -0.03}$ ~pc and $0.14\substack{+0.03 \\ -0.03}$ ~pc for the Plummer p=4 fits, and finally, $0.13\substack{+0.03 \\ -0.03}$ ~pc and $0.14\substack{+0.04 \\ -0.03}$ ~pc for the FWHM derived from the second moment for the northern and southern regions, respectively. 
    \item The width, regardless of the method used,  
    is not correlated with the central column density. Although there is more than an order of magnitude scatter in width, neither sub nor super-critical filament widths are coupled to the column density. 
    The \citet{Heitsch2013b} model with magnetized and externally pressured filaments best reproduces the observed decorrelation.% the best. 
    The model agrees with our observations except for high column densities (10$^{23}$ cm$^{-2}$) where narrower widths are predicted. We found that even at these high column densities traced by our filaments the lack of correlation persists, however, we note that we have a statistically smaller sample at the higher column density end. Accreting filaments in a magnetized environment similar to \citet{Heitsch2013b} model have already been observed in Taurus \citep[e.g.][]{Palmeirim2013} with striations of material perfectly aligned with the magnetic field surrounding the main filament. A similar analysis using, e.g.~BISTRO polarization data \citep{Pattle2017} can be conducted with the CARMA-NRO Orion Survey data in a future study. 
    \item Given the lack of correlation with the column density, we attribute the gradient of the widths all across the cloud to the fact that the filaments are not isolated, homogeneous structures. They are composed of substructure, surrounded by nearby filaments and they form hubs. Connected and disconnected filaments have fluctuating widths.  We find that the majority of filaments closer to the star forming hubs have larger widths. We quantify the complexity of the intensity profiles by looking at the number of shoulders that each profile has. This examination reveals a linear relation between the filament widths and the number of shoulders. From this linear relation we obtain widths of 0.09$\pm$0.02~pc and 0.12$\pm$0.01~pc for a filament with no substructure in northern and southern Orion A, respectively. We suggest that the complexity of the intensity profiles contribute significantly to the spread of the filament width distribution.   
    \item The majority of the identified filaments do not resemble the \textit{fibers} reported in a study towards the Taurus molecular cloud which spatially overlap but are distinct coherent structures in velocity space along the line-of-sight \citep{Hacar2013}. The only fiber-like filaments we find are located in the OMC-4 region where we observe clear multiple velocity components along the line of sight that form coherent structures. The fact that the filaments identified in N$_2$H$^+$ towards the northern ISF by \citet{Hacar2018} and the majority of the filaments identified in this study do not show a spatial overlap implies that the general picture of the filamentary structure in Orion A may differ from what has been observed towards the low mass star forming region Taurus.
 
\end{itemize}

\begin{acknowledgements}
We thank the referee for their insightful comments that helped improve this manuscript and FilChaP.
STS, ASM, PS and VOO acknowledge funding by the Deutsche Forschungsgemeinschaft (DFG) via the Sonderforschungs\-bereich SFB 956 Conditions and Impact of Star Formation (subprojects A4, A6, C1, and C3) and the Bonn-Cologne Graduate School. SDC acknowledges support from the ERC starting grant No. 679852 RADFEEDBACK.
RJS gratefully acknowledges support from an STFC Ernest Rutherford fellowship. This research was carried out in part at the Jet Propulsion Laboratory which is operated for NASA by the California Institute of Technology. PP acknowledges support by the Spanish MINECO under project AYA2017-88754-P (AEI/FEDER,UE). HGA and SK acknowledge support from the National Science Foundation through grant AST-1140063. HGA and SK acknowledge support from the National Science Foundation through grant AST-1140063. \\
\textit{Software:} astropy \citep{astropy2018}, matplotlib \citep{Hunter2007}, scipy \citep{Jones2001}, pandas \citep{pandas2010}, APLpy \citep{aplpy2012}. 
\end{acknowledgements}

\bibliographystyle{aa}
\bibliography{bibliography}

\begin{appendix}
\newpage
\section{\texttt{FilChaP}: Filament Characterization Package}
\label{appendix:filchap}
\texttt{FilChaP} is a python-based software package that can be used to characterize physical properties of filaments. \texttt{FilChaP} derives radial intensity profiles perpendicular to filament spines from which it calculates filament width, skewness and kurtosis. In addition, \texttt{FilChaP} can also calculate filament length and curvature,  as well. The algorithm is publicly available for download at https://github.com/astrosuri/filchap. The download folder includes a python tutorial with an example filament skeleton and a 3D fits cube which the users can utilize to run FilChaP and derive filament properties. The version of \texttt{FilChaP} used to produce the results in this paper has a specific DOI (\url{https://doi.org/10.5281/zenodo.2222325}) and is archived on Zenodo.

\texttt{FilChaP} requires filament coordinates as an input file to derive information on filaments. For this particular study, we have used the output filaments of DisPerSE. These output filaments are simple ASCII files indicating filament ID, x-coordinate (RA), y-coordinate (DEC), z-coordinate (velocity channel). These 4 parameters are given for each point along a filament. If the filaments are identified on a 2-dimensional map instead, then the z-coordinate that corresponds to the velocity channel does not exist in the output file and is not required by \texttt{FilChaP}, and the code can be used in 2D mode. 

The filament width calculation method, as it is the main focus of this study, is explained thoroughly in Section~\ref{sec:width}. Below we explain the baseline removal method used, and present `synthetic' filament networks which we constructed with known widths for the purpose of testing \texttt{FilChaP}.

\subsection{Baseline Subtraction}
\texttt{FilChaP} uses Asymmetric Least Squares Smoothing (AsLS) method for baseline subtraction \citep{Eilers2005}. A commonly used way of achieving a baseline is to smooth the profile until peaks or dips of interest are no longer prominent. The most common smoothing method is the least square method which minimizes the deviations of the smooth counterpart $h$, from the noisy data, $f$. The least square method weighs both negative and positive deviations of $h$ from $f$ the same. In the AsLS method however, deviations from the estimated baseline are weighted asymmetrically, with the positive deviations being weighted less. Because if $f-h$ is a good representation of $f$'s baseline, the deviations should always be positive ($f > h$) due to the fact that $f$ will always have signal (peaks) that take values higher than $h$.

\begin{figure*}
   \centering
   \includegraphics[scale=0.6]{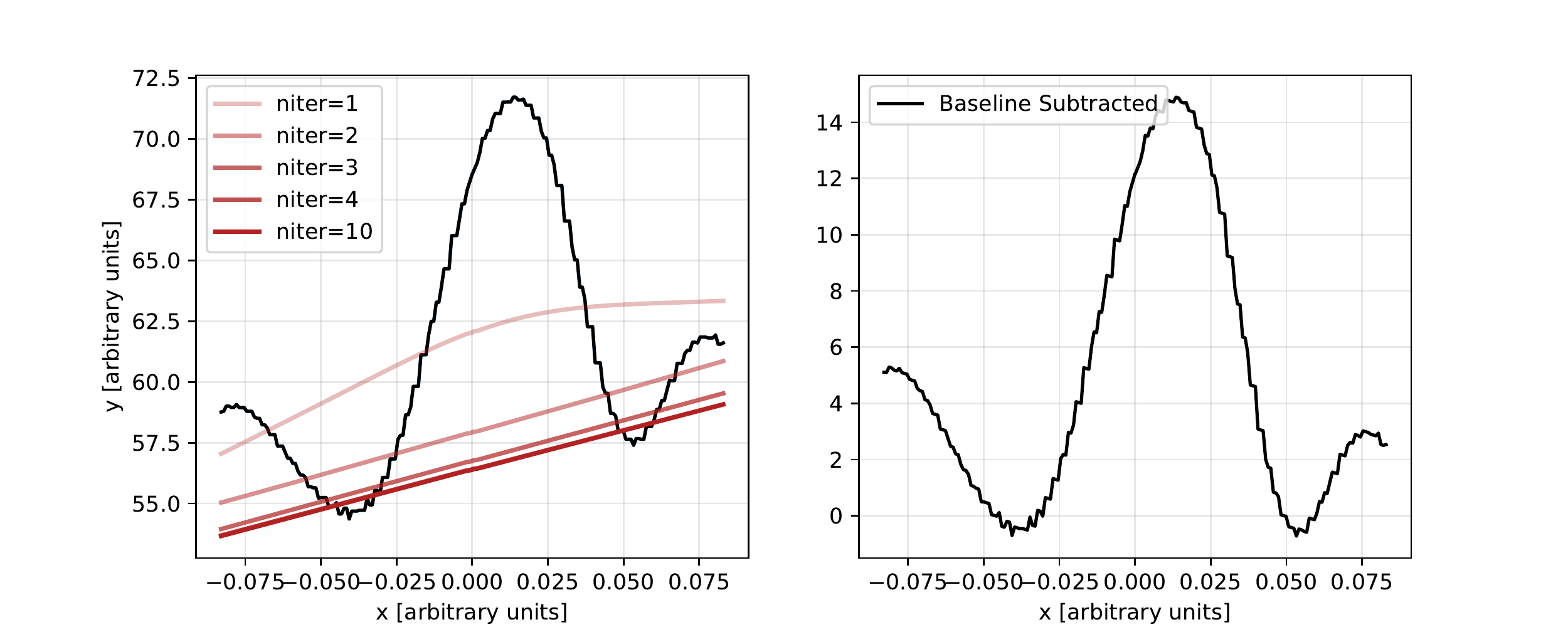} 
      \caption{Example baseline subtraction process. \textit{Left}: Data with a baseline gradient is shown with the black solid curve. The curves with different shades of red represent the AsLS baselines calculated with different number of iterations. The solutions for the baseline converge rather quickly within 10 iterations. \textit{Right}: The baseline subtracted profile.}
        
         \label{fig:baselineSubtraction}
\end{figure*}

Figure~\ref{fig:baselineSubtraction} shows an example of the baseline subtraction process. On the left panel a representative intensity profile with a baseline gradient is shown. This gradient has to be removed before \texttt{FilChaP} can fit a function to calculate its FWHM. The different shades of solid red lines represent the AsLS baselines, $h$, calculated at different number of iterations. After only a few iterations the calculated weights quickly converge, hence a final baseline is obtained. The right panel of the plot shows the baseline subtracted profile. \citet{Eilers2005} gives a \texttt{MATLAB} code for the iterative process of calculating the baselines, and this code is converted to \texttt{python} under \url{https://stackoverflow.com/questions/29156532/python-baseline-correction-library/29185844}. This python code is used by \texttt{FilChaP}. One possible drawback of the AsLS method is that once the baseline is removed a portion of the emission could go below zero. In the future we plan to use \texttt{STATCONT} to prevent this \citep{Sanchez2018}.

\subsection{Tests on the width calculation method}

\begin{figure}
   \centering
   \includegraphics[width=\hsize]{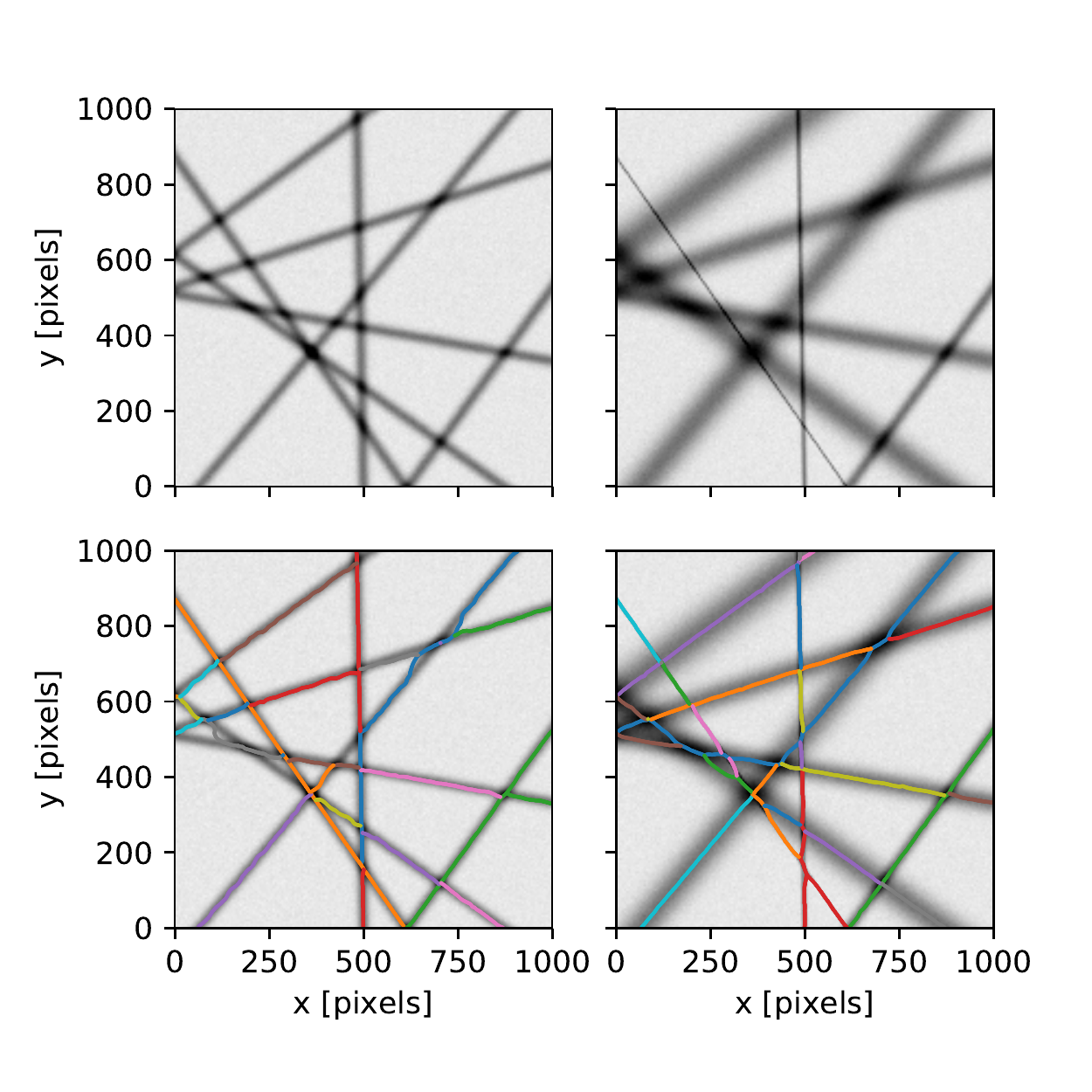} 
      \caption{\textit{Top}: Synthetic filamentary network generated with prefixed filament widths on a $1000\times1000$ pixel grid. On the left, all the filaments have a same width of 25 pixels, while on the right, each filament has a different width that vary from 5 to 85 pixels (see Fig.~\ref{fig:filchap_testCase_FWHM}).
      \textit{Bottom}: Filaments identified by \texttt{DisPerSE}.}
         \label{fig:filchap_testCases}
\end{figure}

\begin{figure}
   \centering
   \includegraphics[width=\hsize]{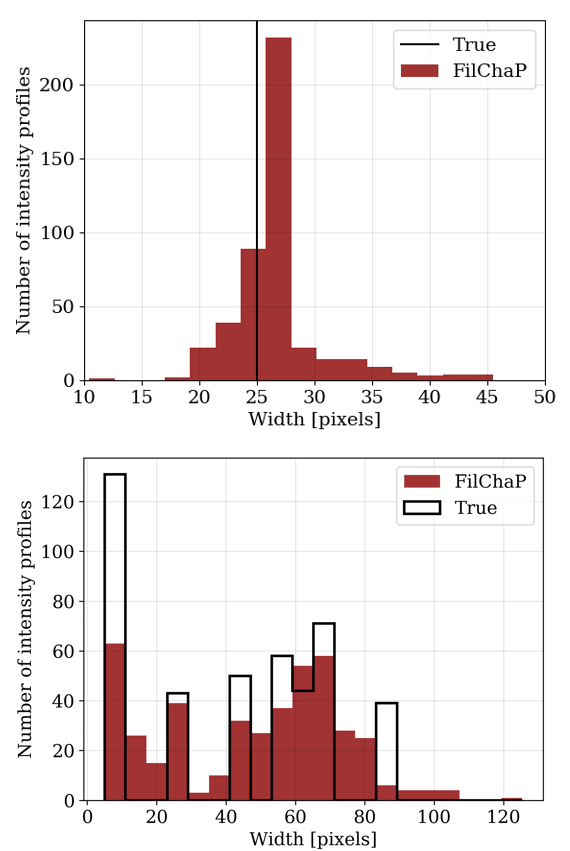} 
      \caption{Distribution of filament widths for the test two cases presented in Figure~\ref{fig:filchap_testCases}. 
      For Case 1 (top panel) the input widths of all the filaments are 25 pixels (solid black line) and 
      the mean of the width distribution calculated with 
      \texttt{FilChap} is 26.5$\pm$4.2 pixels. For Case 2 (bottom panel), the true distribution of the input filament widths is given with the black histogram, and the width calculation that results using \texttt{FilChap} is given with the red histogram. The Spearman correlation coefficient between the two datasets is 0.77.}
         \label{fig:filchap_testCase_FWHM}
\end{figure}

Figure~\ref{fig:filchap_testCases} shows the generated filaments width equal widths (Case 1; top left), and unequal widths (Case 2; top right). The exercise here is to see whether in two different cases, (1) where all filaments have the same width and (2) where all filaments have a different width, \texttt{FilChaP} can recover the width distribution. Both cases have an amount of random Gaussian noise that is selected based on the peak intensity-to-noise ratio in our \c18o integrated intensity map. In Case 1, all the filaments have a 25 pixel width. In the \c18o map, 25 pixels correspond to a 0.1~pc filament width in spatial scales. For Case 2, we allow filament width to vary between 5 and 85 pixels. This is to assess \texttt{FilChaP}'s performance in detecting extreme cases of narrow filaments and diffuse filaments. While for Case 1 we expect to find a distribution that is strongly peaked at around 25 pixels, for Case 2 we expect a wide distribution that has the contribution from all the different scales from 5 pixels to 85 pixels. Figure~\ref{fig:filchap_testCase_FWHM} shows the \texttt{FilChaP} width calculation result for Case 1 (top panel) and Case 2 (bottom panel).  As anticipated, the width distribution for Case 1 has a narrow peak, with a mean of 26.5$\pm$4.2 pixels (where the error corresponds to the 95\% confidence level). For Case 2, we plot both the true input distribution of filament widths and the \texttt{FilChaP} output distribution. In order to quantify the association between the input and output width distributions for Case 2, we calculate Spearman's rank correlation coefficient. This coefficient gives a non-parametric measure of the relation between the two distributions which are connected with a monotonic function. The coefficient takes values between -1 and +1, where the coefficient takes a value of -1 in case of a perfect decreasing monotonic correlation, +1 in case of an increasing monotonic correlation and 0 in case of no correlation. The Spearman's rank coefficient between our the two distributions of Case 2 is 0.77; indicating a strong link between the two datasets. It is clear that the spread of the distribution for Case 1 comes from the errors of individual fits, and the fact that the filaments cross each other at various points throughout the map, which in turn introduces larger widths. For Case 2, in addition to large filament widths at crossing points and the fitting errors, the large spread comes from the underlying dissimilarities between the filament widths. In comparison, the two different cases produce significantly different width distributions that \texttt{FilChaP} successfully captures.

\subsection{Additional functions}
In addition to the filament width, \texttt{FilChaP} calculates filament skewness, kurtosis, length and curvature. The second moment of the intensity profiles is calculated to assess their FWHM, along with the second moment, the third and fourth moments; skewness and kurtosis are also calculated within the width calculation function of \texttt{FilChaP}. The length and curvature are calculated in separate functions, based on the filament spines extracted using DisPerSE. 

\subsubsection{Skewness and kurtosis}
\begin{figure}
   \centering
   \includegraphics[width=0.9\hsize]{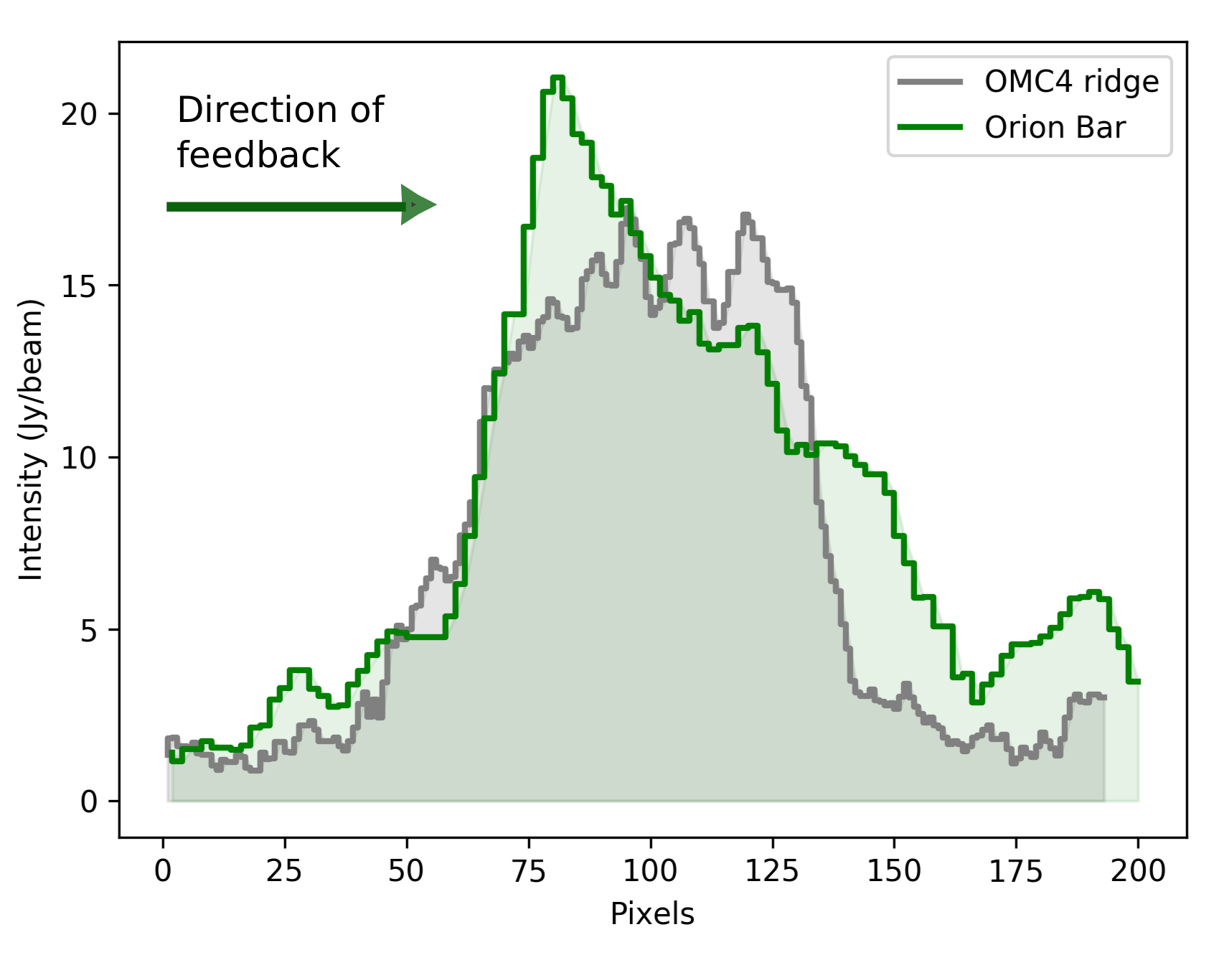} 
      \caption{Comparison of two intensity profiles with different values of skewness and kurtosis. The intensity profile for the Orion Bar position is positive skewed with a tail to the right, while the intensity profile of the OMC4 ridge position is flat in the top resulting in a platykurtic or negative kurtosis distribution. \texttt{FilChaP} determines the skewness and kurtosis of the filament intensity profiles.}
         \label{fig:skewKurt}
   \end{figure}

The skewness and kurtosis provide information about the shape of the intensity profiles. This shaped can be related to physical properties of the filaments and their environments. For example, if a filament is in a feedback region, we expect the intensity profile to be sharper on the side where the filament faces the direction of the feedback. Kurtosis, on the other hand, is a measure of peakiness. Therefore it can indicate locations of high-density, as the CO isotopologues may get optically thick, or in case of \c18o, freeze out that causes the profiles to flatten in the top. Figure~\ref{fig:skewKurt} demonstrates how we can detect the variations of skewness and kurtosis in Orion A. The two plotted intensity profiles are calculated from the \13co emission map and show slices accros the Orion Bar, a well known PDR and OMC-4, the dense star forming region, in green and grey, respectively. The direction of the feedback to the PDR from the Trapezium cluster is shown with a green arrow. The side of the PDR profile towards the feedback forms a sharp edge, whereas the side away from the feedback extends slowly. The OMC-4 ridge shows no such skewness as it is much less affected from the feedback. This plot also shows the difference between the peakiness of both profiles, which can then be quantified with the kurtosis parameter. The following paper on comparing filament properties in different tracers (Suri et al.~in prep) touches upon the PDR nature of the filaments in feedback regions and compares properties of intensity profiles using skewness and kurtosis analysis.  
 
\subsubsection{Length and curvature}
In order to obtain the filament length, \texttt{FilChaP} calculates the \textit{projected} distances between the consecutive skeleton points along a filament. This is a projected distance, because in observational datasets we do not have a knowledge on the spatial 3rd dimension of the filaments. In addition, the code offers calculating ``kinkiness'', or in other words; curvature, of filaments by comparing the calculated length of the filament to the distance between the start and end points. Therefore, we define curvature as $k = r/R$ where $r$ is the distance between the end points and $R$ is the calculated length of the filament. A $k$ value of unity indicates a straight filament.

\begin{figure}
   \centering
   \includegraphics[width=0.9\hsize]{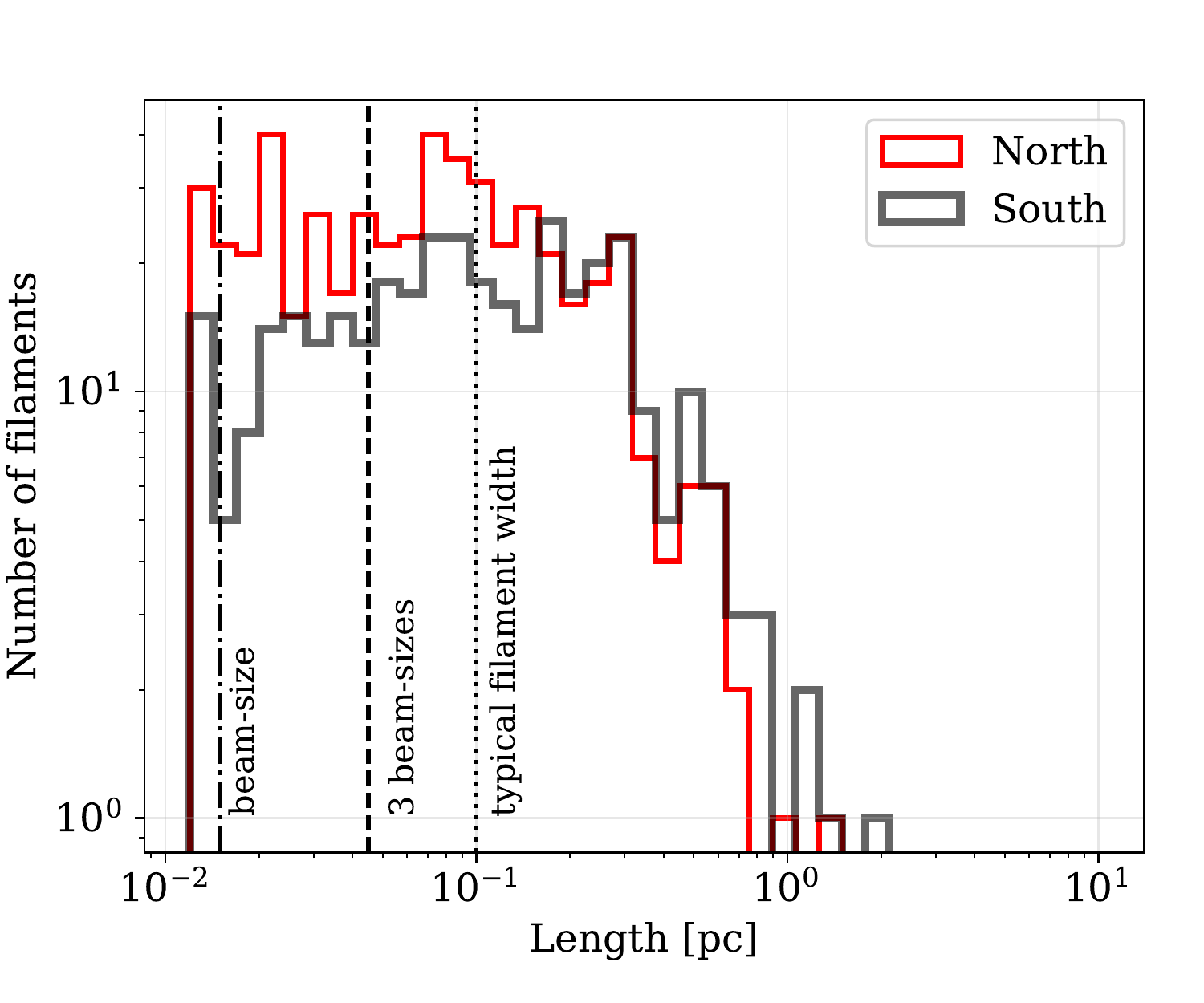} 
      \caption{Distribution of filament lengths in the north (red) and south (black). The dot-dashed, dashed and the dotted lines respectively indicate a beam-size, 3 beam-sizes and the typical filament width found in this study.}
         \label{fig:lenghtKDE}
   \end{figure}

These two properties of filaments; length and curvature, are highly dependent on the parameters used to identify filaments with DisPerSE. Therefore, we note that the resulting statistics should be interpreted with caution. However, the credibility of these parameters can be improved in the future by testing different filament identification methods, comparing identified filamentary structures to one another, and using multiple tracers to assess the connection. Nevertheless, for the identified filaments in \c18o using DisPerSE with the set of parameters explained in Section\ref{sec:filamentProperties}, we measured filament length between 0.03-1.8 pc. Figure~\ref{fig:lenghtKDE} shows the distribution of the resulting filament lengths for the northern (red) and southern Orion A (black). The dashed vertical line at 0.045~pc indicates 3-beamsizes, and for the width calculations we discard the structures that are shorter than 0.045~pc.  

\begin{figure}
   \centering
   \includegraphics[width=0.9\hsize]{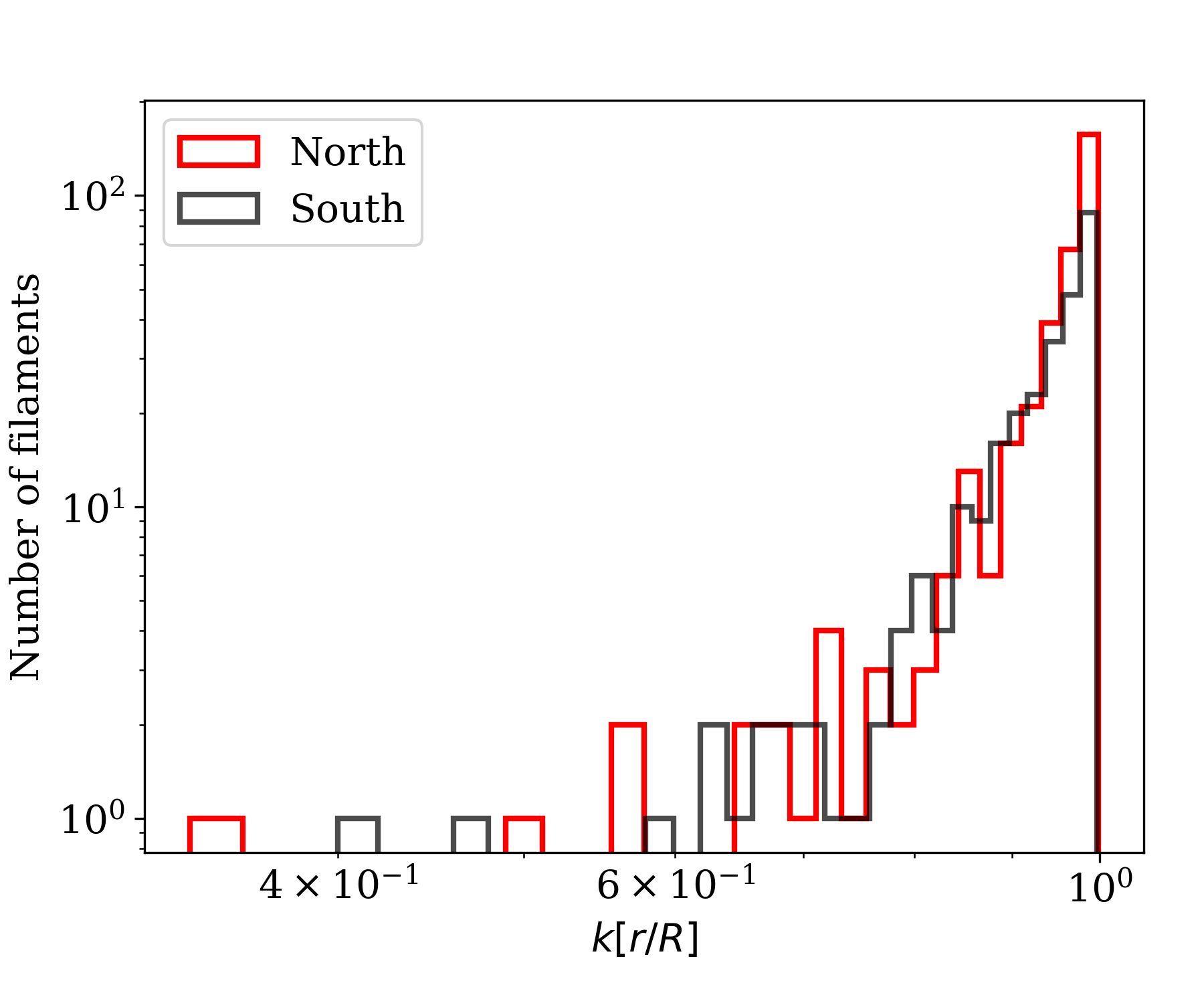}
      \caption{Distribution of filament kinkiness, $k$, where $k = r/R$ with $r$ being the distance between the end points and $R$ the calculated length of the filament. The majority of the filaments in Orion~A are found to be unbent.}
         \label{fig:curvature}
   \end{figure}

The curvature values of filaments in Orion A north and south are shown in histograms in Figure~\ref{fig:curvature}. 
There are a few filaments that appear to be more bent, and have $k$ values of between 0.3--0.9. 
The most bent filament in the north is located in OMC4 ridge. Majority of filaments, however, have very low curvature.

\section{Additional Material}\label{appendix:additional}
In Section~\ref{sec:filamentProperties}, we presented width distributions for all the identified filaments that are longer than 3 beam-sizes and are fitted with a reduced chi-squared less than 6. Here, we reproduce Figures~\ref{fig:width_results_all}, \ref{fig:peaksVSwidth} and \ref{fig:widthVScolumnDensity} considering different criteria on filament and profile selection. We look at the width distributions of (1) all filaments and intensity profiles fitted with a reduced chi-squared less than or equal to 2, (2) filaments with aspect ratios larger than 2 and intensity profiles that are fitted with a reduced chi-squared less than or equal to 6, (3) filaments with aspect ratios larger than 2 and intensity profiles fitted with a reduced chi-squared less than or equal to 2. This different selection criteria are aimed to ensure that we do not introduce any bias in the determined widths when including all the filaments that are longer than 3 beam-sizes and have a reduced chi-squared less than 6.

\begin{figure}%[t]
   \centering
   \includegraphics[width=\hsize]{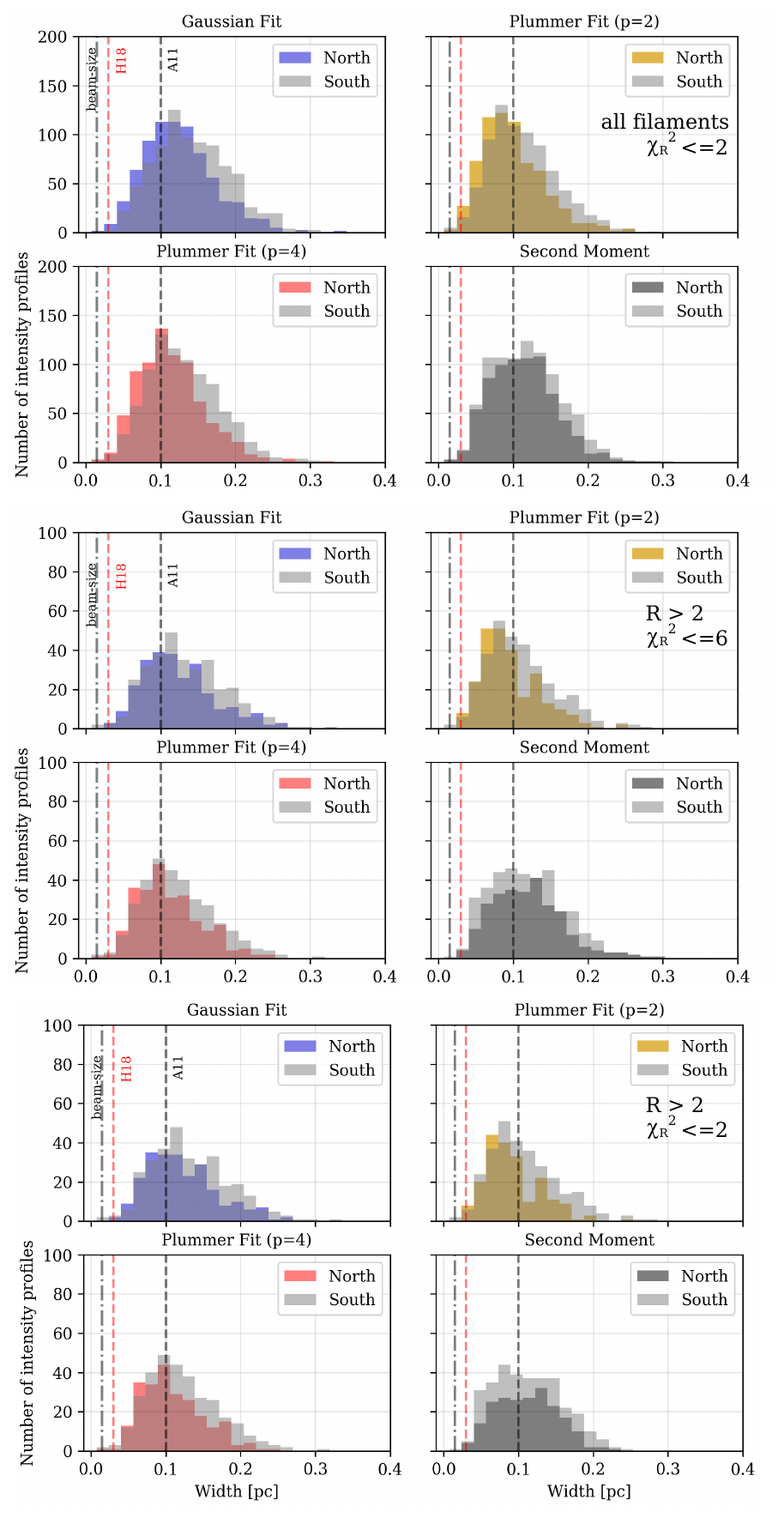}
      \caption{Width distributions for four different width calculation methods (Gaussian, Plummer-like with p=2 and p=4 and second moment, as shown in Fig.~\ref{fig:width_results_all}) when the filament aspect ratio and the reduced chi-squared criteria are varied. The \textit{top four panels} show the distributions of all filaments and intensity profiles fitted with a reduced chi-squared less than or equal to 2. The \textit{middle four panels} show the width distributions for filaments with aspect ratios larger than 2 and fits with chi-squared less than or equal to 6. Finally, the \textit{bottom four panels} show the width distributions for filaments with aspect ratios larger than 2 and reduced chi-squared values less than or equal to 2.}
         \label{fig:app_width_dist}
\end{figure}

\begin{figure}%[t]
   \centering
   \includegraphics[width=\hsize]{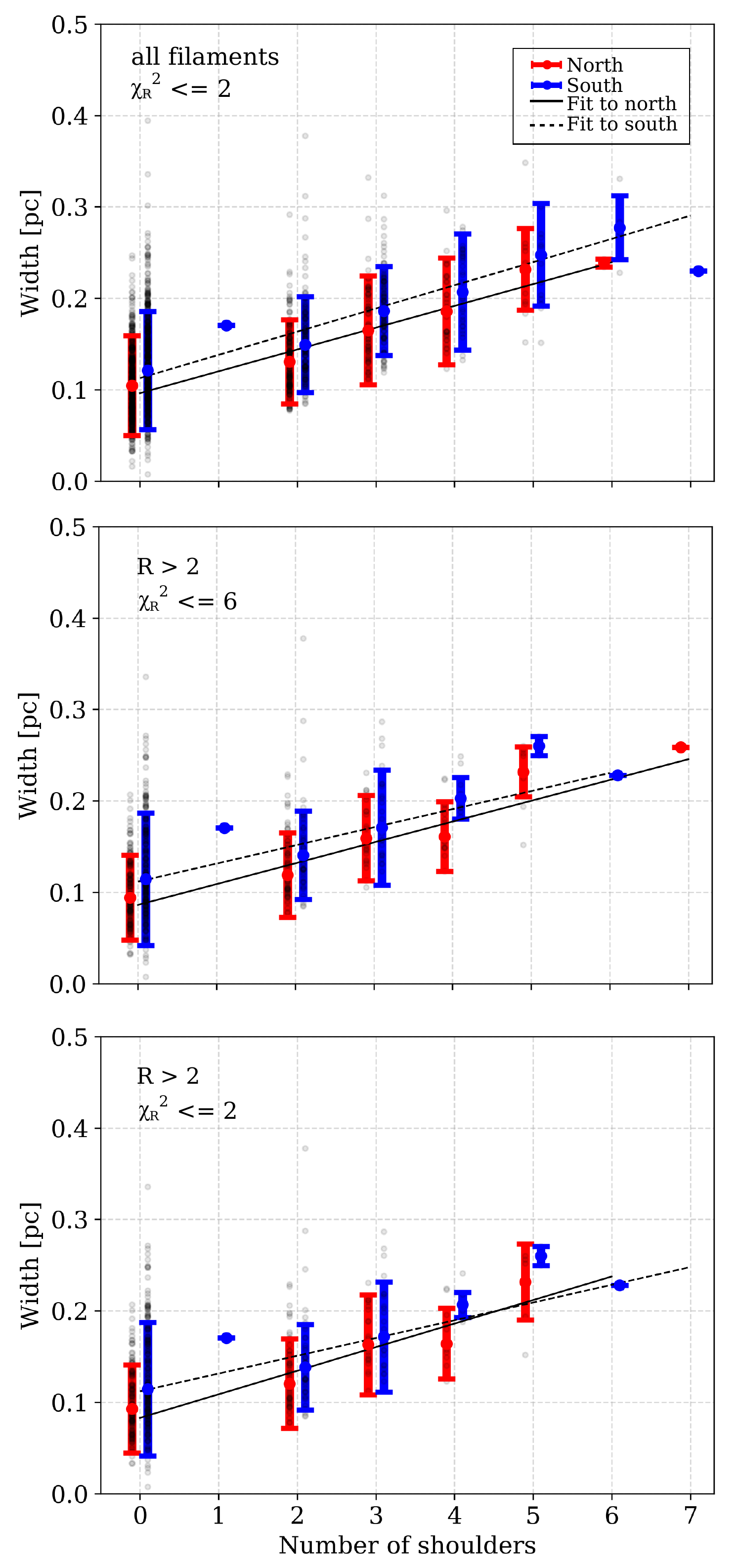}
      \caption{Same as Figure~\ref{fig:peaksVSwidth} for the different criteria (aspect ratio and reduced chi-squared of the fit) studied in Section~\ref{appendix:additional}.}
         \label{fig:app_width_shoulders}
\end{figure}

\begin{figure}%[t]
   \centering
   \includegraphics[width=\hsize]{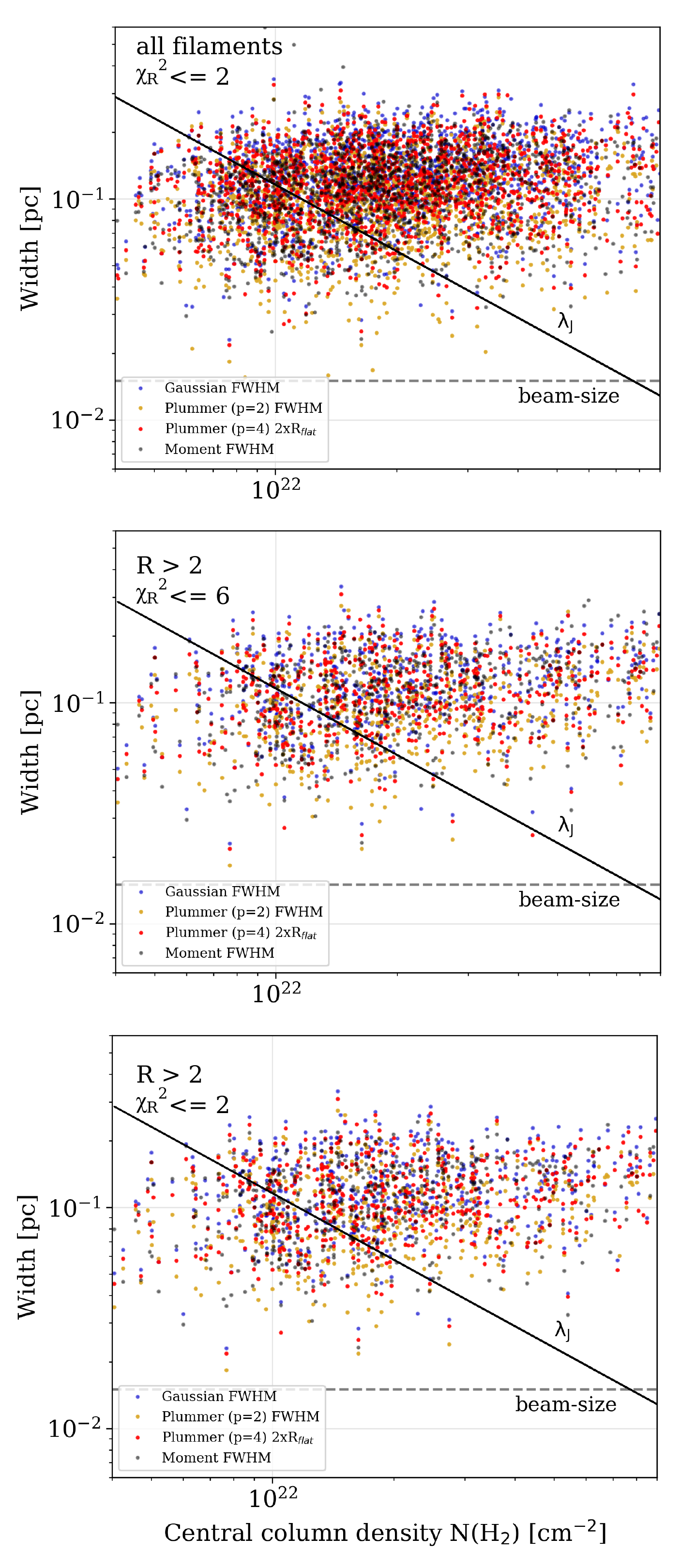}
      \caption{Same as Figure~\ref{fig:widthVScolumnDensity} for the different criteria (aspect ratio and reduced chi-squared of the fit) studied in Section~\ref{appendix:additional}.}
         \label{fig:app_width_density}
\end{figure}

\begin{table*}[ht]
\caption{Median filament widths according to different selecting criteria (see Figure~\ref{fig:app_width_dist}), and inferred filament width when no shoulders are identified in the intensity profile (intercept of the linear fits of Figure~\ref{fig:app_width_shoulders}).}              % title of Table
\label{table:1}      % is used to refer this table in the te
\centering                                      % used for centering table
\begin{tabular}{c c c c c c}          % centered columns (4 columns)
\hline\hline                        % inserts double horizontal lines
\multirow{2}{*}{Criteria} & Gaussian & Plummer (p=2) & Plummer (p=4) & Second Moment & Intercept \Tstrut \\
 & north/south & north/south & north/south & north/south & north/south\Bstrut \\
% table heading
\hline                                   % inserts single horizontal line
    \multirow{2}{*}{all, ${\chi}_R^2\leq$ 6}
    & $0.12\substack{+0.15 \\ -0.09}$~pc & $0.09\substack{+0.13 \\ -0.07}$~pc & $0.11\substack{+0.14 \\ -0.08}$~pc & $0.12\substack{+0.15 \\ -0.09}$~pc & 0.09$\pm$0.02~pc \Tstrut\Bstrut \\      
    & $0.14\substack{+0.18 \\ -0.11}$~pc & $0.11\substack{+0.14 \\ -0.08}$~pc & $0.13\substack{+0.16 \\ -0.09}$~pc & $0.12\substack{+0.15 \\ -0.09}$~pc & 0.11$\pm$0.01~pc \Tstrut\Bstrut \\
\hline
     \multirow{2}{*}{all, ${\chi}_R^2\leq$ 2}
     & $0.12\substack{+0.15 \\ -0.09}$~pc & $0.09\substack{+0.12 \\ -0.07}$~pc & $0.11\substack{+0.14 \\ -0.08}$~pc & $0.11\substack{+0.14 \\ -0.08}$~pc & 0.10$\pm$0.01~pc \Tstrut\Bstrut\\
     & $0.14\substack{+0.18 \\ -0.11}$~pc & $0.11\substack{+0.14 \\ -0.08}$~pc & $0.12\substack{+0.16 \\ -0.10}$~pc & $0.12\substack{+0.15 \\ -0.08}$~pc & 0.11$\pm$0.01~pc \Tstrut\Bstrut\\
\hline
    \multirow{2}{*}{ R > 2, ${\chi}_R^2\leq$ 6}
    & $0.12\substack{+0.15 \\ -0.09}$~pc & $0.09\substack{+0.12 \\ -0.07}$~pc & $0.10\substack{+0.14 \\ -0.08}$~pc & $0.12\substack{+0.15 \\ -0.09}$~pc & 0.09$\pm$0.02~pc \Tstrut\Bstrut\\
    & $0.13\substack{+0.17 \\ -0.09}$~pc & $0.10\substack{+0.14 \\ -0.08}$~pc & $0.12\substack{+0.15 \\ -0.09}$~pc & $0.11\substack{+0.15 \\ -0.08}$~pc & 0.11$\pm$0.01~pc \Tstrut\Bstrut\\
\hline
    \multirow{2}{*}{ R > 2, ${\chi}_R^2\leq$ 2}
    & $0.11\substack{+0.15 \\ -0.09}$~pc & $0.08\substack{+0.12 \\ -0.06}$~pc & $0.10\substack{+0.14 \\ -0.08}$~pc & $0.11\substack{+0.15 \\ -0.08}$~pc & 0.08$\pm$0.03~pc \Tstrut\Bstrut\\
    & $0.12\substack{+0.17 \\ -0.09}$~pc & $0.10\substack{+0.13 \\ -0.07}$~pc & $0.12\substack{+0.15 \\ -0.08}$~pc & $0.11\substack{+0.15 \\ -0.08}$~pc & 0.11$\pm$0.01~pc \Tstrut\Bstrut\\

\hline                                             %inserts single line
\end{tabular}
\end{table*}

The width distributions for the cases mentioned above are shown in Figure~\ref{fig:app_width_dist} top, middle and bottom panels, respectively. For the majority of the cases, regardless of the criteria, the southern region of Orion~A has filaments with larger widths as found in Fig.~\ref{fig:width_results_all}. The distribution of Plummer-like (p=2) fits results in the most narrow median filament width ($0.08\substack{+0.12 \\ -0.06}$~pc) when only those filaments with an aspect ratio larger than 2 and a reduced chi-squared value less than 2 is selected. The median filament widths for these three cases in the northern and southern regions are listed in Table~\ref{table:1}. The median widths in all cases are in the range 0.08~pc to 0.14~pc, similar to the range found when considering all the filaments.

In Fig.~\ref{fig:app_width_shoulders}, we reproduce the correlation between the filament width and the number of shoulders for the varying selection criteria considered above (i.e., aspect ratio and reduced chi-squared). Similar linear fits are found in all cases. The inferred filament width when no shoulders are identified in the intensity profiles for each of the cases considered is listed in the last column of Table~\ref{table:1}.

Finally, in Fig.~\ref{fig:app_width_density} we investigate that the no correlation found between filament widths and H$_2$ column density is not biased by the selection of the filaments. The bottom panel in Fig.~\ref{fig:app_width_density} is the most strict selection of filaments (filaments with aspect ratio larger than 2 and reduced chi-squared less than 5) suggest no correlation between width and column density as discussed when considering all the filaments.

\end{appendix}

\end{document}